\RequirePackage{fix-cm}
\documentclass[twocolumn,epjc3]{svjour3}  
\smartqed  % flush right qed marks, e.g. at end of proof
\RequirePackage{graphicx}
\journalname{Eur. Phys. J. A}

\usepackage{graphicx}% Include figure files
\usepackage{dcolumn}% Align table columns on decimal point
\usepackage{bm}% bold math
\usepackage{color}
\usepackage{bbold}
\usepackage{ulem}

\usepackage{hyperref}

\def\bra#1{\langle#1\vert}
\def\ket#1{\vert#1\rangle}

\usepackage{amsmath}
\usepackage{amssymb}
\usepackage{url}

\def\bmDelta{\mbox{\boldmath$\Delta$}}

\def\simle{\mathrel{\rlap{\raise 0.511ex \hbox{$<$}}{\lower 0.511ex \hbox{$\sim$}}}}
\def\simge{\mathrel{ \rlap{\raise 0.511ex \hbox{$>$}}{\lower 0.511ex \hbox{$\sim$}}}}

\newcommand{\rmd}{{\rm d}}
\newcommand{\rme}{{\rm e}}

\newcommand \beq{\begin{eqnarray}}
\newcommand \eeq{\end{eqnarray}} 

\def\bea{\begin{eqnarray}}
\def\eea{\end{eqnarray}}
\def\be{\begin{equation}}
\def\ee{\end{equation}}

\def\simle{\mathrel{\rlap{\raise 0.511ex \hbox{$<$}}{\lower 0.511ex 
\hbox{$\sim$}}}}
\def\simge{\mathrel{ \rlap{\raise 0.511ex 
\hbox{$>$}}{\lower 0.511ex \hbox{$\sim$}}}}

\newcommand{\nn}{\nonumber\\ }

\def\x{{\boldsymbol x}}

\def\b{{\boldsymbol b}}
\def\x{{\boldsymbol x}}

\def\r{{\boldsymbol r}}
\def\p{{\boldsymbol p}}

\begin{document}

\title{Angular structure of many-body correlations in atomic nuclei}
\subtitle{From nuclear deformations to diffractive vector meson production in $\gamma A$ collisions}

\author{Jean-Paul Blaizot\thanksref{e1,addr1}
        \and
        Giuliano Giacalone\thanks{CERN-TH preprint number: CERN-TH-2025-057}\thanksref{e2,addr2}
}

\thankstext{e1}{jean-paul.blaizot@ipht.fr}
\thankstext{e2}{giuliano.giacalone@cern.ch}

\institute{Institut de Physique Th\'eorique, Universit\'e Paris Saclay, CEA, CNRS, F-91191 Gif-sur-Yvette, France \label{addr1}
           \and
           Theoretical Physics Department, CERN, CH-1211 Gen\`eve 23, Switzerland \label{addr2}
}

\date{Received: date / Accepted: date}

\maketitle

\begin{abstract}
 There is growing evidence that high-energy scattering processes involving nuclei can offer unique insights into the many-body correlations present in nuclear ground states, in particular those of \textit{deformed} nuclei. These processes involve, for instance, the collective anisotropic flows in heavy-ion collisions, or the diffractive production of vector mesons in photo-nuclear ($\gamma A$) interactions.  In this paper, we use a classical approximation and simple analytical models in order to exhibit characteristic and universal features of ground-state correlation functions that result from the presence of a deformed intrinsic state. In the case of a small axial quadrupole deformation, we show that the random rotation of the intrinsic density of the nucleus leads to a specific quadrupole modulation of the lab-frame two-body density as a function of the relative azimuthal angle. As a phenomenological, albeit academic application, we analyze the diffractive production of vector mesons in high-energy $\gamma+^8$Be collisions. This demonstrates with the simplest deformed nucleus how the two-body correlations impact the $|t|$ dependence of the incoherent cross sections.
\end{abstract}

\tableofcontents

%%%%%%%%%%%%%%%%%%%%%%%%%%%%%%%%%%%%%%%%%%%%%%%%%%%%%%
%%%%%%%%%%%%%%%%%%%%%%%%%%%%%%%%%%%%%%%%%%%%%%%%%%%%%%
%%%%%%%%%%%%%%%%%%%%%%%%%%%%%%%%%%%%%%%%%%%%%%%%%%%%%%
%%%%%%%%%%%%%%%%%%%%%%%%%%%%%%%%%%%%%%%%%%%%%%%%%%%%%%
%%%%%%%%%%%%%%%%%%%%%%%%%%%%%%%%%%%%%%%%%%%%%%%%%%%%%%
%%%%%%%%%%%%%%%%%%%%%%%%%%%%%%%%%%%%%%%%%%%%%%%%%%%%%%

\section{Introduction}
\label{sec:1}

Atomic nuclei are among the most fascinating and complex systems in nature, exhibiting emergent collective phenomena that manifest across a wide spectrum of scales \cite{Nazarewicz:2016gyu}. Most strikingly, many nuclei appear to be ``deformed'' into various shapes \cite{bohr}, reflecting spatial correlations of nucleons which play an essential role in explaining experimental data on nuclear structure \cite{Heyde:2016sop}.

For decades, deformations and collectivity in atomic nuclei have been explored through low-energy experimental techniques, such as Coulomb excitation \cite{Cline:1986ik} or laser spectroscopy \cite{Yang:2022wbl}, which have allowed the characterization of the electromagnetic properties of thousands of nuclear states. In recent years, high-energy scattering experiments have emerged as a new tool for probing nuclear deformations \cite{Jia:2022ozr}. Quite remarkably, collective flow measurements in high-energy nuclear collisions reveal a rather strong dependence on the low-energy structure of the colliding ions~\cite{STAR:2015mki,ALICE:2018lao,CMS:2019cyz,ATLAS:2019dct,STAR:2021mii,ALICE:2021gxt,ATLAS:2022dov,STAR:2024wgy,ALICE:2024nqd}, showing sensitivity to their shapes \cite{Xu:2021uar,Nijs:2021clz,Zhang:2021kxj,Zhao:2022uhl,Ryssens:2023fkv,Fortier:2023xxy,Giacalone:2024luz,Xu:2024bdh,Fortier:2024yxs,Giacalone:2024ixe,Mantysaari:2024uwn}, their skin thickness \cite{Li:2018oec,Xu:2021vpn,Giacalone:2023cet,Liu:2023pav}, and potential clustering phenomena \cite{YuanyuanWang:2024sgp,Zhao:2024feh,Prasad:2024ahm,Wang:2024ulq,Liu:2025zsi}. Similarly, deep inelastic scattering in $\gamma A$ collisions is also sensitive to the nuclear structure \cite{STAR:2022wfe,Mantysaari:2023qsq,Mantysaari:2023prg,Lin:2024mnj}. In all these processes at high energy, the characteristic time scale involved in the interaction is much shorter than that of the intrinsic nuclear dynamics. Thus, compared to the electromagnetic probes mentioned above, these processes occur on time scales short enough to probe the instantaneous distributions of nucleons in the ground state, offering a most \textit{direct} view of the nucleon positions and their correlations \cite{Giacalone:2023hwk,Duguet:2025hwi}.

Motivated by this perspective, this work seeks to advance our understanding of nuclear deformation by studying its impact on generic, model-independent features of the many-body correlations associated with the ground states of deformed nuclei.  Specifically, we use a classical approximation for the correlation functions, which amounts to assuming that the correlations are generated from the random rotation of an intrinsic deformed state \cite{Jia:2021tzt,Jia:2021qyu}, in line with the rotor picture of nuclear rotations. This allows us to analytically characterize some of the dominant features of angular correlations. In particular, for small quadrupole deformations, we show that the density-density correlation function $\langle \rho(\r_1)\rho(\r_2)\rangle$ behaves as $\delta^2 r_1^2r_2^2\cos (2\varphi_{12})$, where $\delta$ is a measure of the deformation, $\r_i=(r_i,\varphi_i)$, $i=1,2$, and $\varphi_{12}=\varphi_1-\varphi_2$. We expect the quadrupole modulation in the relative azimuthal angle, $\varphi_{12}$, with peaks at $\varphi_{12}=0$ and $\varphi_{12}=\pi$, with an amplitude that grows as one moves toward the edges of the nuclear system, to be universal (for axially symmetric systems with stable ground-state quadrupole deformation). It emerges naturally in simple analytical models and we expect that it will be present in more sophisticated treatments of nuclear correlations, i.e. going beyond the classical approximation used here. 

As an illustration of an observable consequence of these two-body correlations, we consider the coherent and incoherent diffractive production of vector mesons in high-energy $\gamma A$ scattering. This process probes the density-density correlation function of the nucleus in its ground state \cite{Caldwell:2010zza}, making it a powerful diagnostic tool. We argue that the impact of the nuclear deformation is maximal for values of the momentum transfer, $|t|$, in the region where the coherent cross section exhibits a diffractive minimum, which should improve the chances of detecting the effect in future experiments \cite{Kesler:2025ksf}.

The manuscript is organized as follows. In Section~\ref{sec:2}, we briefly review the concept of nuclear deformation, and that of intrinsic state. We consider nuclei with a stable deformation and use a classical approximation to infer the general form of the many-body ground state correlation functions associated to the rotation of the intrinsics state, emphasizing in particular the role of the density-density correlation function. Section~\ref{sec:3} presents explicit analytical calculations of this density-density correlation function in different models of the intrinsic state: a simple rotating thin rod, a dumbbell, and a Gaussian density mimicking a deformed nucleus. In Section~\ref{sec:4}, the same analysis is repeated for a realistic simple nucleus, $^{8}$Be, described in  a harmonic-oscillator model. This admittedly somewhat academic example is used to calculate the coherent and incoherent production of vector mesons in high-energy $\gamma$+$^{8}$Be collisions, and to demonstrate how nuclear deformation can be extracted from the corresponding cross sections. Section~\ref{sec:5} is left for concluding remarks.

%%%%%%%%%%%%%%%%%%%%%%%%%%%%%%%%%%%%%%%%%%%%%%%%%%%%%%
%%%%%%%%%%%%%%%%%%%%%%%%%%%%%%%%%%%%%%%%%%%%%%%%%%%%%%
%%%%%%%%%%%%%%%%%%%%%%%%%%%%%%%%%%%%%%%%%%%%%%%%%%%%%%
%%%%%%%%%%%%%%%%%%%%%%%%%%%%%%%%%%%%%%%%%%%%%%%%%%%%%%
%%%%%%%%%%%%%%%%%%%%%%%%%%%%%%%%%%%%%%%%%%%%%%%%%%%%%%
%%%%%%%%%%%%%%%%%%%%%%%%%%%%%%%%%%%%%%%%%%%%%%%%%%%%%%

\section{Correlations induced by rotations}
\label{sec:2}

When we talk about deformed nuclei we have in mind nuclei whose shape deviates from the spherical shape. For instance, there exist many axially symmetric nuclei, with a stable ground state deformation that can be characterized by the quadrupole moment of the mass distribution
\beq
Q=\sum_{i=1}^A \left(2z_i^2-x_i^2-y_i^2\right)=\sum_{i=1}^A 2\,r_i^2\, P_2(\cos\theta_i),
\eeq
where $P_2$ is a Legendre polynomial. Here, we have chosen the $z$ axis to be the symmetry axis, and $A$ is the total number of nucleons. However, it turns out that the ground states of most of such nuclei are eigenstates of the total angular momentum $J$, with $J=0$, i.e., they are invariant under rotation. The ground-state expectation value of the quadrupole operator, which depends on an orientation, therefore vanishes. In fact, the relation between the deformation and the properties of the ground state is a subtle one. The usual interpretation of the deformed state is that of an intrinsic state whose collective rotation allows the nucleus to acquire good angular momentum \cite{bohr}. Thus, in the simplest models, like rotor models,  a rotationally-invariant state emerges from an angular average over all orientations of the intrinsic state.

Since the square of the quadrupole moment operator contains a scalar component, the expectation value of the operator $Q^2$ in the exact ground state wave function needs not vanish. This expectation value is then often used to get quantitative information about the deformation of a nucleus (see e.g. \cite{Poves:2019byh}). The ground-state expectation value of $Q^2$ can be written thus 
\beq
\bra{\Psi_0}Q^2\ket{\Psi_0}=\sum_{n\ne 0} \bra{\Psi_0} Q\ket{\Psi_n}\bra{\Psi_n} Q\ket{\Psi_0},
\eeq
 where $\ket{\Psi_0}$ denotes the ground state and  $\ket{\Psi_{n\ne 0}}$ the excited states. For nuclei with a stable deformation, and exhibiting a rotational spectrum, the sum over intermediate states is often dominated by the first excited states with angular momentum $J=2$. In the case where only the contribution of the lowest state is significant, we have approximately 
\beq
\bra{\Psi_0}Q^2\ket{\Psi_0}\approx \left|\bra{\Psi_0} Q\ket{\Psi_2}\right|^2, 
\eeq
the latter quantity being proportional to the so-called $B(E2; 0^+ \rightarrow 2^+)$ which controls the strength of electromagnetic transitions carrying angular momentum $2$. Such matrix elements have been studied for a long time in nuclear physics, as they are accessible in various measurements \cite{Raman:2001nnq,Pritychenko:2013gwa}. 

At this point, we note that since $Q$ is a local one-body operator: 
\beq
Q=\sum_{i=1}^A 2r_i^2 P_2(\cos\theta_i)=\sum_{i=1}^A \hat q(\r_i)=\int_\r \,\hat\rho(\r) q(\r),\eeq
with $\hat\rho(\r)$ the density operator
\beq 
\hat\rho(\r)=\sum_{i=1}^A \delta(\r-\r_i),
\eeq
the expectation value $\langle Q^2 \rangle $ can be written as follows
\beq\label{eq:Q2rhorho}
\langle Q^2 \rangle =\!\int_{\r_1 \r_2} q(\r_1) q(\r_2)\,\langle \hat\rho(\r_1)\hat \rho(\r_2)\rangle,
\eeq
where the angular brackets denote the average over the ground state of the nucleus, and we used $\int_\r$ as a shorthand for $\int\rmd^3 \r$. The quantity $\langle \hat\rho(\r_1)\hat \rho(\r_2)\rangle$ is part of the  density-density correlation function, defined by 
\beq\label{eq:Sxydef}
S(\r_1,\r_2)&=&\langle \hat\rho(\r_1)\hat \rho(\r_2)\rangle-\langle \hat\rho(\r_1)\rangle\langle \hat \rho(\r_2)\rangle.
\eeq
and indeed since  $\int_{\r} q(\r) \,\langle \hat\rho(\r)\rangle=0$, one could rewrite Eq.~(\ref{eq:Q2rhorho}) as
\beq
\langle Q^2 \rangle =\!\int_{\r_1 \r_2} q(\r_1) q(\r_2)\,S(\r_1,\r_2),
\eeq
As we shall see in Section~\ref{sec:4} the same density correlation function is directly probed in the diffractive photo-production of vector mesons on nuclei, the Fourier transform of $S$ being proportional to the incoherent cross section of the process.\footnote{In low energy electron scattering experiments, a more general object enters, namely $S(\r_1,t_1,\r_2,t_2)$ where the dependence on $t_1-t_2$ carries information on the internal dynamics of the nucleus, for instance on the energies of its excited states. At high energy, the interaction time is so small in comparison with the time scales that characterize the intrinsic nuclear dynamics that only the instantaneous correlation function is probed.} It is then legitimate to broaden the discussion and ask whether one can characterize the specific correlations that are induced by the rotation of a deformed nucleus. In the case of local two-body operators, such as $Q^2$, the relevant correlations are contained in the density-density correlation function, as we have just argued. Our main goal in this paper is to characterize these correlations and identify their origin.\footnote{We have focused here on the quadrupole deformation, but the discussion can be trivially generalized to the case of higher multipole moments of the density, such as $Q_{lm}=\sum_{i=1}^A r_i^ll Y_l^m(\theta_i,\varphi_i)$. The important point in the present discussion is that $Q_{lm}$ is a one-body operator, and that the correlations are captured by the density-density correlation function. Obviously, other operators may play important roles in the study of nuclear deformations, such as $Q^4$, $Q^6$, etc. \cite{Kumar:1972zza}, which involve correlation functions of higher order. These could be analyzed following the same lines as those presented in this paper for the two-point functions.}

We are interested in measurements that probe the nucleus on very short time scales. The outcome of such measurements is then determined, to a large extent, by the spatial configuration of the nucleons at the instant of the measurement. This configuration fluctuates event by event, the probability $P(\r_1,\cdots,\r_A)$ that the nucleons occupy the positions $\r_1,\cdots,\r_A$ at the time of measurement being given by 
\beq\label{eq:proba1toA}
P(\r_1,\cdots,\r_A)=\left| \bra{\r_1,\cdots,\r_A} \Psi_0\rangle  \right|^2,
\eeq
where $\ket{\Psi_0}$ is the (rotationally-invariant) ground state wave function. In fact, we are interested in high energy processes where colliding particles are moving on straight line trajectories which are not deflected during the collisions. Following Glauber's theory and the eikonal approximation  \cite{Miller:2007ri,Loizides:2014vua,dEnterria:2020dwq}, most observables are then determined by the integral of $P(\r_1,\cdots,\r_A)$ along the direction of the collision axis, which we choose to be the $z$ axis. Denoting by $\b_i$ the coordinates in the transverse plane, we then define
\beq\label{eq:Pbidef}
P(\b_1,\cdots,\b_A)=\int_{z_1\cdots z_A}\left| \bra{\b_1,z_1\cdots,\b_A,z_A} \Psi_0\rangle  \right|^2,
\eeq
where $\int_z$ is a shorthand for $\int\rmd z$. We similarly define the density operator in the transverse plane 
\beq \hat t(\b)= \int_z\sum_{i=1}^A\delta(\r-\hat{\r}_i)= \sum_{i=1}^A \delta(\b-\hat{\b}_i).\eeq

Given the probability $P(\b_1,\cdots,\b_A)$, we define the one-point function  $t^{(1)}(\b)$  and  the two-point function $t^{(2)}(\b,\b')$ in the usual way
\beq\label{eq:1ptdef}
\langle \hat t(\b)\rangle=At^{(1)}(\b),
\eeq
and
\beq\label{eq:2ptdef}
\langle \hat t(\b)\hat t(\b')\rangle=At^{(1)}(\b)\delta(\b-\b')+ A(A-1)t^{(2)}(\b,\b'),\nn
\eeq
where here, for any function $f(\b_1\cdots \b_A)$,
\beq
\langle f\rangle=\int_{\b_1\cdots \b_A} \,P(\b_1,\cdots,\b_A) f(\b_1\cdots \b_A).
 \eeq
Note that the one- and two-point functions are normalized according to
\beq
\int\rmd^2 \b \,t^{(1)}(\b)=1=\int\rmd^2 \b \int\rmd^2 \b'\, t^{(2)}(\b,\b').  
\eeq
In terms of these functions, the density-density correlation reads
\beq\label{eq:Sxydef2}
&&S(\b,\b')= At^{(1)}(\b) \delta(\b-\b') +A(A-1) t^{(2)}(\b,\b') \nn &&\qquad\qquad\qquad\qquad\qquad\qquad -A^2 t^{(1)}(\b)t^{(1)}(\b'),
\eeq
where the first line in Eq.~(\ref{eq:Sxydef2}) corresponds to the first term of Eq.~(\ref{eq:Sxydef}). 
Note also the properties 
\beq\label{eq:sumruleS}
\int_{\b, \b^\prime}  S(\b,\b')=0, \quad  t^{(1)}(\b)= \int_{\b^\prime}  t^{(2)}(\b,\b'),
\eeq
which follow from the definitions above.

We now turn to the question of characterizing the dominant qualitative features of the angular correlations associated with the collective rotation of deformed nuclei. Such correlations are of classical nature, being associated with a rotation, and to identify their main characteristic, we shall rely on rotor models, such as those presented in the next section. However, we expect that the results extracted from classical considerations are generic and would also emerge in more sophisticated microscopic treatments.

\begin{figure}[t]
    \centering
    \includegraphics[width=0.8\linewidth]{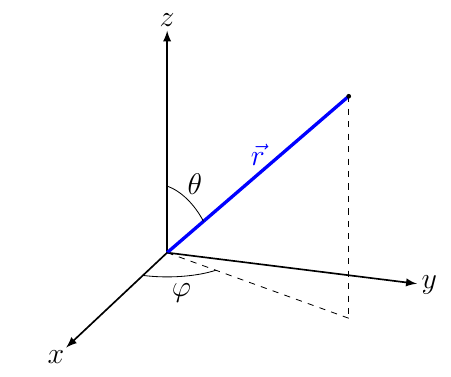}
    \caption{Spherical coordinate system, $\vec r =(r,\theta,\varphi)$, adopted in this work.}
    \label{fig:coords}
\end{figure}

Since we consider nuclei with a stable deformation, in a measurement taking place on a time scale much shorter than the period of the rotational motion, the orientation of the nucleus is well defined. Let us denote by $\Omega=(\theta,\varphi)$ the two angles that specify the orientation of the symmetry axis of the nucleus in the laboratory frame (see Fig.~\ref{fig:coords} for the definition of the angles). We then define  
\beq \label{eq:POmega}
P_\Omega(\r_1,\cdots,\r_A)=|\bra{\r_1,\cdots,\r_A}\Phi_ \Omega\rangle|^2,
\eeq
where $\ket{\Phi_ \Omega}$ denotes the intrinsic state. We interpret $P_\Omega(\r_1,\cdots,\r_A)$ as the probability of finding a configuration $\r_1,\cdots,\r_A$ of nucleon positions, given the orientation $\Omega$ of the intrinsic state. Since a rotor model assumes rigid rotation, the dependence on $\Omega$ of $P_\Omega$ is easily obtained. The coordinates $\r_1,\r_2,\cdots,\r_A $ in Eq.~(\ref{eq:POmega}) are measured in the laboratory frame. When the body-fixed axis coincides with the laboratory axis, $\Omega=0$. For a general orientation $\Omega$, we have   
\beq
P_\Omega(\r_1,\r_2,\cdots,\r_A)=P_{\Omega=0}(\r_1',\r_2',\cdots,\r_A'), 
\eeq
where  $\r_i'={\cal R}_\Omega^{-1} \r_i$, with ${\cal R}_\Omega$ the rotation  which aligns the symmetry axis of the nucleus in the direction specified by the angles $\theta$ and $\varphi$. This relation extends to all $n$-body densities associated to $P_\Omega$. For instance, the rotated intrinsic density reads
\beq\label{eq:rotateddensity}
\rho_\Omega(\r)=\rho_{\Omega=0}({\cal R}_\Omega^{-1} \r),
\eeq
where $\rho_{\Omega=0}$ is the density in the body-fixed frame and $x,y,z\; (=\r)$ are the laboratory coordinates.

By integrating $P_\Omega(\{\r_i\})$ over all the $z_i$ variables, as in Eq.~(\ref{eq:Pbidef}) above, we obtain the probability $P_\Omega(\{\b_i\})$ of a configuration of nucleons in the transverse plane, for a given orientation $\Omega$ of the nucleus.  
Our first major assumption, consistent with the rotor model, is that 
\beq \label{eq:approx1}
P(\b_1,\cdots,\b_A)=\int\frac{\rmd\Omega}{4\pi}P_\Omega(\b_1,\cdots,\b_A).
\eeq
This entails that the (lab-frame) one and two-point functions can be written as 
\beq\label{eq:approx2}
&&t^{(1)}(\b)=\int\frac{\rmd\Omega}{4\pi}\,t^{(1)}_\Omega(\b),\nn && t^{(2)}(\b,\b')=\int\frac{\rmd\Omega}{4\pi}\,t^{(2)}_\Omega(\b,\b'),
\eeq
where $t_\Omega^{(1)}(\b)$ and $t^{(2)}_\Omega(\b,\b')$ can be deduced from (\ref{eq:1ptdef}) and (\ref{eq:2ptdef}), respectively, after substituting $P_\Omega$ to $P$. 
Thus, for instance, 
\beq
\langle \hat t(\b)\hat t(\b')\rangle_\Omega=At_\Omega^{(1)}(\b)\delta(\b-\b')+ A(A-1)t_\Omega^{(2)}(\b,\b')\nn
\eeq
where $\langle\cdots\rangle_\Omega$ denotes an average at fixed $\Omega$ (i.e. with probability $P_\Omega$).

Our second major approximation is to assume that, in the intrinsic state, the nucleons are uncorrelated\footnote{We ignore short range correlations, as well as those due to Pauli exclusion principle, which we do not expect to have a major impact on angular correlations. These are known for instance to have modest effects on angular correlations and spatial anisotropies in heavy-ion collisions \cite{Alvioli:2011sk,Blaizot:2014wba,Alvioli:2018jls,Zhang:2024vkh}. We shall also ignore later (in Section \ref{sec:4}) the correlations due to the center of mass motion, although these could be sizable for light nuclei.}, so that 
\beq\label{eq:approx2}
t^{(2)}_\Omega(\b,\b')=t^{(1)}_\Omega(\b)t^{(1)}_\Omega(\b').
\eeq
Thus, the only correlations are those that develop as a result of the angular average, in which
\beq
\int\frac{\rmd\Omega}{4\pi}t^{(1)}_\Omega(\b)t^{(1)}_\Omega(\b')\ne \int\frac{\rmd\Omega}{4\pi}t^{(1)}_\Omega(\b)\;\int\frac{\rmd\Omega'}{4\pi}t^{(1)}_{\Omega'}(\b').\nn
\eeq
Note that such angular correlations affect similarly all $n$-point functions with arbitrary $n$. 

  In the next section, we focus on these angular averages, and it is convenient to give a specific name to the corresponding correlation functions. Thus we define 
  \beq
  G(\b,\b')=\int\frac{\rmd\Omega}{4\pi}t^{(1)}_\Omega(\b)t^{(1)}_\Omega(\b'),
  \eeq
  and its connected part
  \beq
   G_c(\b,\b')=G(\b,\b')-\int\frac{\rmd\Omega}{4\pi}t^{(1)}_\Omega(\b)\;\int\frac{\rmd\Omega'}{4\pi}t^{(1)}_{\Omega'}(\b').
   \eeq
These functions are normalized thus
\beq
\int_\b G(\b,\b')=t^{(1)}(\b'),\quad \int_{\b,\b'} G(\b,\b')=1.
\eeq

The equations (\ref{eq:approx1}) and (\ref{eq:approx2}) completely specify our calculations. The angular average is akin to an average over events. In each event one probes a particular configuration of nucleons which corresponds to some global orientation of the nucleus (even if this orientation is not accurately defined in each event). The averaging over events thus includes the averaging over these orientations, which produces the angular correlations. This strategy coincides with that used in nearly all analysis of flow phenomena in heavy-ion experiments \cite{Giacalone:2023hwk}.

Observables in scattering experiments most often involve the Fourier transform of the correlation functions, rather than their coordinate dependence. Thus, if $\bmDelta$ denotes the momentum transfer in a collision of the nucleus with a probe, the relevant Fourier transform of the density correlation function is
\beq\label{eq:SDelta}
S(\Delta)=\int_{\b_1, \b_2}  \,\rme^{-i\bmDelta\cdot(\b_1-\b_2)} S(\b_1,\b_2).
\eeq
We define similarly the Fourier transform of the thickness function, $\langle t(\Delta)\rangle$, and of the correlation function, $G(\Delta)$.  Note that, because of rotational invariance, these Fourier transforms depend only on the modulus of $\Delta=|\bmDelta|$. Note also that, as a consequence of the first Eq.~(\ref{eq:sumruleS}), $S(\Delta=0)=0$.  We anticipate that correlations exist within a spatial region on the order of the size, $R$, of the nucleus, that is, for $R\Delta\lesssim 1$. For $R\Delta\gg 1$, correlations vanish. However, $S(\Delta)$ contains a contact term that represents local density fluctuations. Taking the Fourier transform of Eq.~(\ref{eq:Sxydef2}) we indeed obtain
\beq \label{eq:SDelta2}
S(\Delta)&=&A+A(A-1) t^{(2)}(\Delta)-A^2 t^{(1)}(\Delta)^2
\eeq
Owing to the approximations (\ref{eq:approx2}), this can be written as 
\beq \label{eq:SDelta3}
S(\Delta)=A-A G(\Delta)+A^2 G_c(\Delta).
\eeq
Since $G(\Delta=0)=\langle t(\Delta=0)\rangle=1$,  one verifies that $S(\Delta=0)=0$. Furthermore,  the constant term dominates for $R\Delta \gg 1$. In a scattering process, this term represents the scattering on $A$ individual uncorrelated point-like nucleons (see Sect.~\ref{sec:4}).  

%%%%%%%%%%%%%%%%%%%%%%%%%%%%%%%%%%%%%%%%%%%%%%%%%%%%%%
%%%%%%%%%%%%%%%%%%%%%%%%%%%%%%%%%%%%%%%%%%%%%%%%%%%%%%
%%%%%%%%%%%%%%%%%%%%%%%%%%%%%%%%%%%%%%%%%%%%%%%%%%%%%%
%%%%%%%%%%%%%%%%%%%%%%%%%%%%%%%%%%%%%%%%%%%%%%%%%%%%%%
%%%%%%%%%%%%%%%%%%%%%%%%%%%%%%%%%%%%%%%%%%%%%%%%%%%%%%
%%%%%%%%%%%%%%%%%%%%%%%%%%%%%%%%%%%%%%%%%%%%%%%%%%%%%%

\section{Rotor models}
\label{sec:3}
The goal of this section is to characterize the main qualitative features of the density-density correlation function of a rigidly rotating axially-symmetric body. To do so, we shall consider various rotor models, gradually approaching a realistic picture of a deformed nucleus.

In this section, all densities will be normalized to unity, so that $t(\b)=t^{(1)}(\b)$. Also, in order to avoid conflicts in the notation, we shall no longer denote the generic transverse coordinates by $\b$ but by $\r$, the three-dimensional coordinates being then $(\r,z)$. We denote by $\Omega=(\theta,\phi)$ the angles that specify the orientation of the symmetry axis of the system, while $\r=(x,y)$, or $\r=(r,\varphi)$ will denote respectively the cartesian and polar transverse coordinates of an arbitrary point in the system.   

\subsection{The thin rod}

\begin{figure}[t]
    \centering
    \includegraphics[width=.9\linewidth]{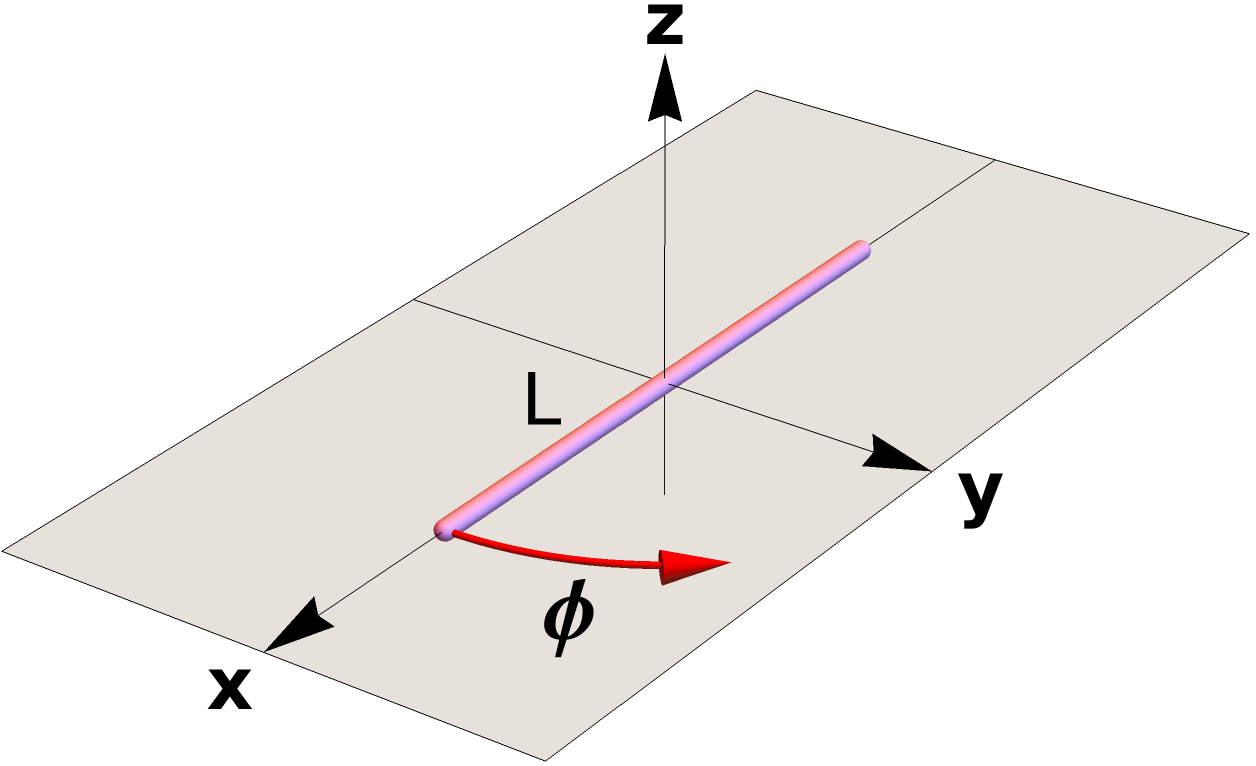}
    \caption{The thin rod rotating in the two-dimensional plane.}
    \label{fig:rod}
\end{figure}

The simplest model illustrating the emergence of angular correlations is that of a thin rod of length $2L$ that rotates around its middle point. The density of the rod is supposed to be uniform. Its orientation in the laboratory frame is determined by the two angles $\theta$ and $\phi$.  When projected onto the plane $(x,y)$, the rod has length $L\sin\theta$. When $\phi=0$ the rod is aligned along the $x$ axis (see Fig.~\ref{fig:rod}). The thickness function is then given by
\beq
t_\theta(x,y)=\frac{1}{2L \sin\theta}\,\Theta(L\sin\theta-|x|)\delta(y).
\eeq
Note that the dependence on $\theta$ amounts simply to the substitution $L\mapsto L\sin\theta$. The averaging over $\theta$ corresponds then to an average over rods of different length. This average  yields a smooth variation with $x$ of the density, 
\beq
\langle t(x)\rangle=\frac{1}{2L}\left( \frac{\pi}{2}-\arcsin\frac{|x|}{L}  \right)
\eeq
which has little (qualitative) consequence for the angular correlations that we are after. We shall then ignore from there on the $\theta$ dependence and fix $\theta=\pi/2$. 

Consider then a rod rotating in the plane $(x,y)$, and oriented with an angle $\phi$ with respect to the $x$-axis (Fig.~\ref{fig:rod}). The thickness is given by $t_\phi(\r)= t_{\phi=0}({\cal R}_\phi^{-1} \r)$ (see Eq.~(\ref{eq:rotateddensity})), that is, with $\r=(x,y)$, 
\beq
t_\phi(x,y)=\frac{1}{2L}\Theta(L-r) \,\delta(-x\sin\phi+y \cos\phi),
\eeq
or,  setting
$
x=r\cos\varphi$, $ y=r\sin\varphi,
$
\beq 
t_\phi(r,\varphi)=\frac{1}{ 2L r} [\delta(\phi-\varphi)+\delta(\phi-\varphi-\pi] \,\Theta(L-r),
\eeq
whose angular average is simply $\langle t(r)\rangle=1/2\pi$ fm$^{-3}$.

Consider now the correlation function ($\r_i=(r_i,\varphi_i))$
 \beq\label{eq:2ptdumbbell}
&&G(\r_1,\r_2)=\int_0^{2\pi} \frac{\rmd\phi}{2\pi} t_\phi(r_1,\varphi_1)t_\phi(r_2,\varphi_2)\nn  &&=\frac{1}{4L^2 r_1 r_2} \int_0^{2\pi} \left[\frac{\rmd\phi}{2\pi}\delta(\phi-\varphi_1)\delta(\phi-\varphi_2-\pi)+\cdots\right] \nn
&&= \frac{1}{2\pi L^2 r_1 r_2} [ \delta(\varphi_{12}) +\delta(\varphi_{12}-\pi)],
\eeq
where $\varphi_{12}\equiv \varphi_1-\varphi_2$, and it is understood that $r_1\le L$, $r_2\le L$. In the first line, we have exhibited only the first of four equivalent terms.  The angle average two-point function exhibits two peaks, one at $\varphi_{12}=0$ and one at $\varphi_{12}=\pi$. This formula clearly illustrates the origin of the correlations: for two points $\r_1$ and $\r_2$ to be correlated, i.e., to contribute to the integral (\ref{eq:2ptdumbbell}), they must belong to the ``same rod". That is, there must be an orientation $\phi$ of the rod such that $(r_1,\varphi_1)$ and $(r_2,\varphi_2)$  can be found on that rod. As we shall see, this sharp angular correlation, producing peaks in the density correlation function at relative angles $\varphi_{12} =0$ and $\varphi_{12}=\pi$,   survives, albeit somewhat smoothened, in the following examples.
\begin{figure}[t]
    \centering
    \includegraphics[width=\linewidth]{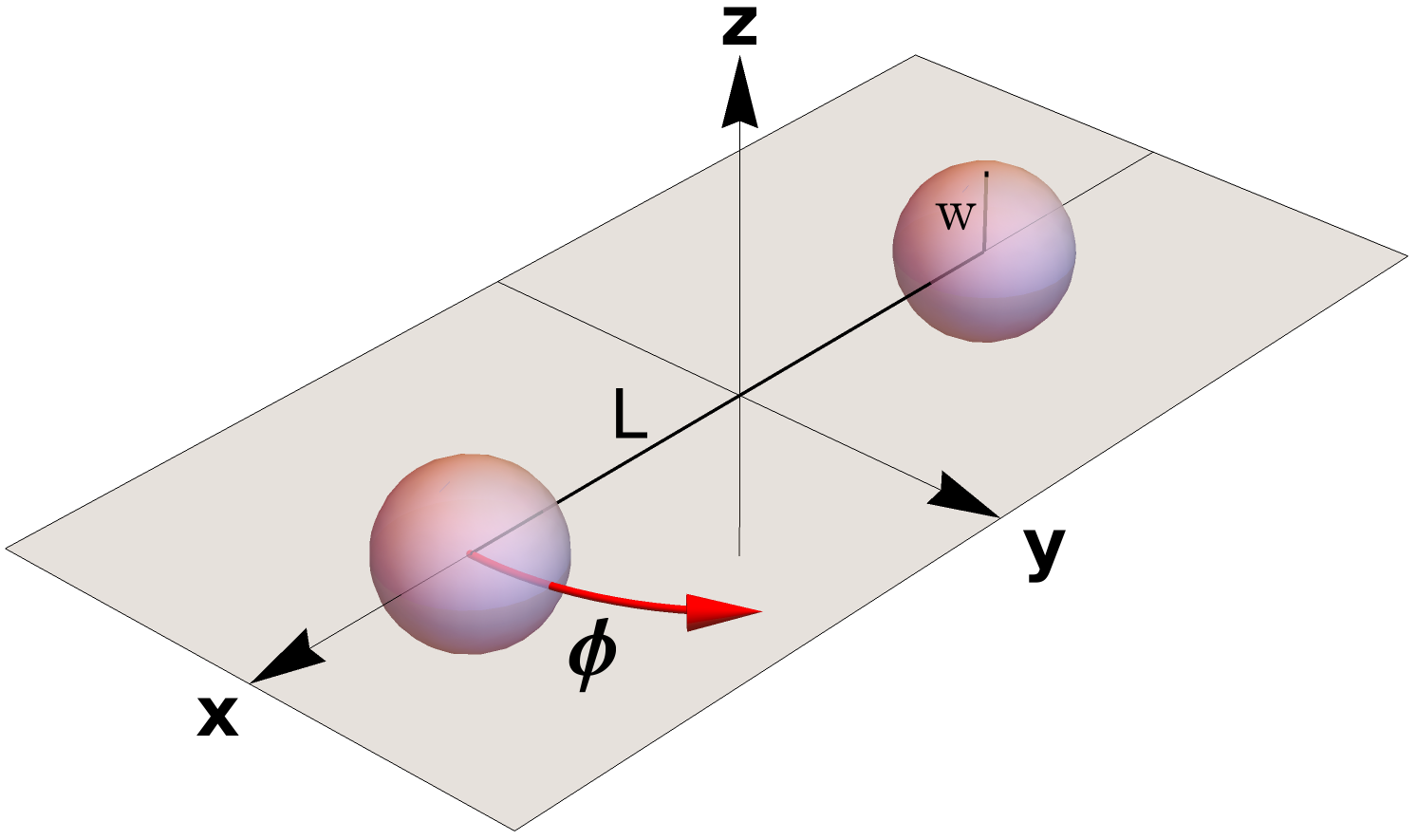}
    \caption{The dumbbell rotating in the plane $(x,y)$ of the laboratory frame.}
    \label{fig:dumbbell}
\end{figure}

\subsection{The dumbbell}

An example richer than that of the rod is a dumbbell, that is, two masses that rotate rigidly around the barycenter. We consider such a dumbbell, made of two spherical blobs centered respectively at positions $+L$ and $-L$ of the body-fixed frame $z$-axis. Each blob has a Gaussian shape with a width $w$. An illustration is given in Fig.~\ref{fig:dumbbell}. The corresponding density reads
\beq
\rho(\r,z)\!=\!\frac{1}{(2\pi)^{3/2}} \frac{1}{2w^3}\!\left[\rme^{-\frac{x^2+y^2+(z-L)^2}{2w^2}} \!\!+\! \rme^{-\frac{x^2+y^2+(z+L)^2}{2w^2}}  \right],\nn
\eeq
and is normalized such that $\int_{\r,z} \rho(\r,z)=1$. In the following, we take $L=1$, in arbitrary units.

\subsubsection{Thickness function}

The thickness function $t_\Omega(x,y)$ is obtained by integrating the rotated density $\rho_\Omega(\r)$ (see Eq.~(\ref{eq:rotateddensity})) with respect to $z$:
\beq\label{eq:thicknessdef}
t_\Omega(x,y)=\int_{-\infty}^\infty \rmd z \, \rho_\Omega(x,y,z).
\eeq
The expectation values of $x^2$ and $y^2$ with respect to the thickness $t_\Omega(x,y)$ are given by  $\langle x^2\rangle =2(L^2\sin^2\theta+w^2)$, and $\langle y^2\rangle=2 w^2$, so that 
\beq\label{eq:defparam}
\varepsilon=\frac{\langle x^2-y^2\rangle}{\langle x^2+y^2\rangle}=\frac{L^2\sin^2\theta/w^2}{2+L^2\sin^2\theta/w^2}.
\eeq
The quantity $0\le \varepsilon\le 1$ measures the eccentricity of the mass distribution projected on the plane orthogonal to the laboratory $z$-axis. It is directly related to the deformation of the system, which we may also characterize by the ratio $w/L$. When $w/L\ll 1$, the deformation is large. In fact, when $w\to 0$, $\varepsilon\to 1$ and the dumbbell becomes equivalent to the thin rod discussed in the previous section. On the other hand, when $w/L\gg 1$, the two blobs overlap and the system becomes nearly spherical. The deformation is then small, and $\varepsilon\simeq \frac{L^2\sin^2\theta}{2w^2}$.

Keeping the angle $\theta$ fixed and averaging over $\phi$, one gets, after a simple calculation, the following expression of the (angle averaged) thickness function
\beq\label{eq:thickdumb}
\langle t_\theta(r)\rangle= \frac{   \rme^{  -\frac{r^2+L^2\sin^2\theta}{2w^2}  }}{2\pi w^2}I_0\left( \frac{Lr\sin\theta}{w^2} \right),
\eeq
where $I_0$ is a modified Bessel function. 
Note that the angle $\theta$ enters the expression of $\langle t_\theta(r)\rangle$ just through the combination $L\sin\theta$. This is a general feature, and changing $\theta$ simply corresponds to changing the apparent length of the system in the plane $(x,y)$. From now on, we shall therefore set $\theta=\pi/2$ and focus on a dumbbell rotating in the transverse plane, as shown in Fig.~\ref{fig:dumbbell}. The effect of averaging over the angle $\theta$ will be considered in a different example, studied in the next section. 

\subsubsection{Correlation function}
\begin{figure}[t]
\begin{center}
\includegraphics[width=.97\linewidth]{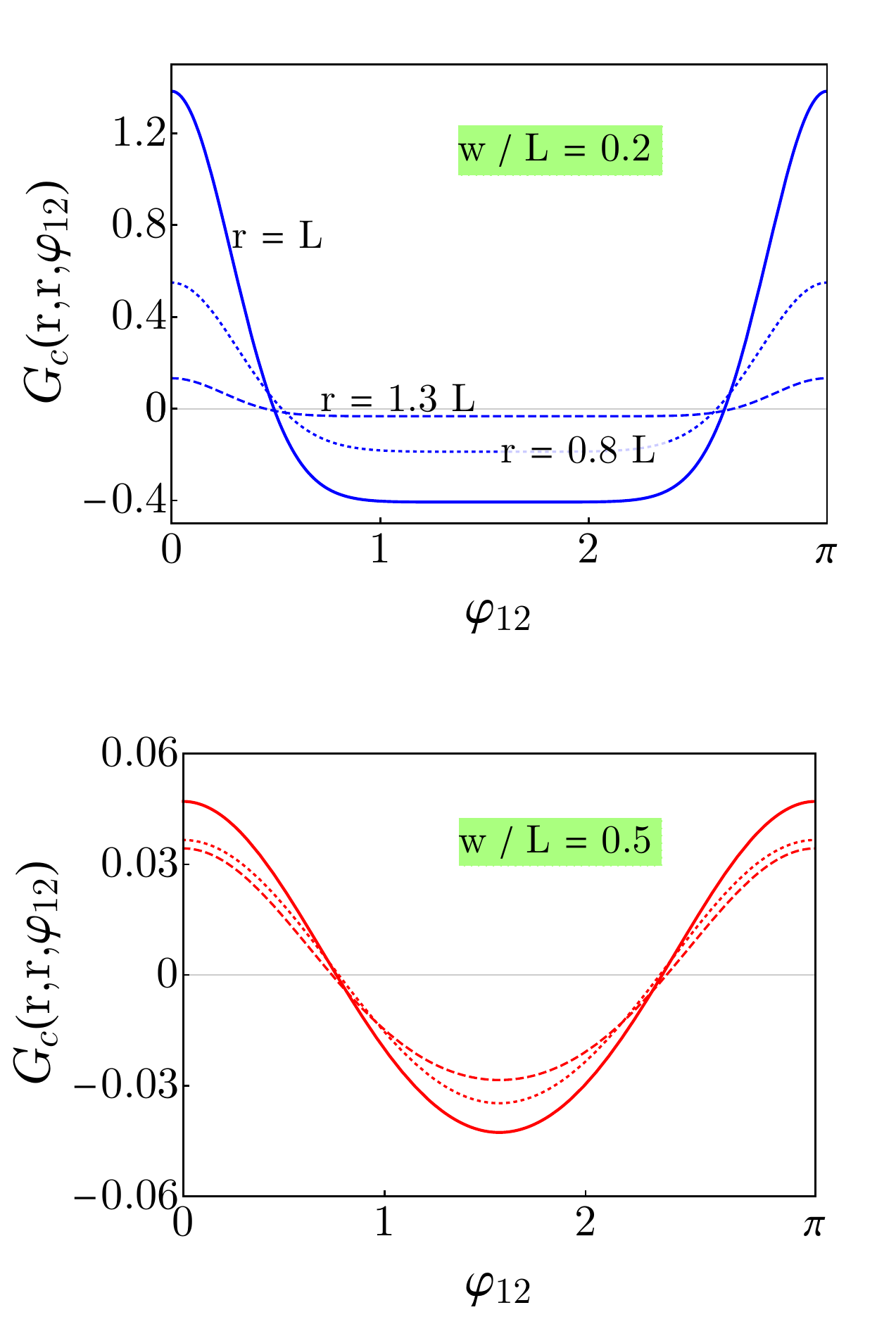}
\caption{The correlation function $G_c(r,r,\varphi_{12})$ as a function of $\varphi_{12}=\varphi_1-\varphi_2$. The dependence on $r$ is indicated by the various line styles: thick for $r=1$, dashed for $r=1.3 L$, dotted for $r=0.8 L$. The angular correlation is clearly visible and gets attenuated as $w$ increases from $w=0.2 L$ (top) to $w=0.5L$ (bottom).}
\label{fig:Gc_dumbbell}
\end{center}
\end{figure}

We now turn to the correlation function and consider first the two-point function $G(\r_1,\r_2)=\langle t(\r_1) t(\r_2)\rangle$, where the angular brackets denote the average over the angle $\phi$. A simple calculation yields 
\beq
&&G(\r_1,\r_2)=\frac{ \rme^{-\frac{2L^2+r_1^2+r_2^2}{2 w^2}} }{8 \pi ^2w^4}\nn &&\qquad\times \left[ I_0\left(\frac{L}{w^2}|\r_1+\r_2|  \right)+ I_0\left(\frac{L}{w^2}|\r_1-\r_2|  \right) \right],
\eeq
with 
\beq
|\r_1\pm \r_2|=\sqrt{r_1^2+r_2^2\pm2r_1 r_2\cos\varphi_{12}}.
\eeq
Here we have set $\r_1=r_1(\cos\varphi_1,\sin\varphi_1)$, and similarly for $\r_2$. Note that, after the average over the orientation of the dumbbell,  the correlation function depends only on the relative angle, $\varphi_{12}=\varphi_1-\varphi_2$ between $\r_1$ and $\r_2$.

The connected correlation function is  given by
\beq\label{eq:Sr1r2alpha12}
&&G_c(\r_1,\r_2)=\frac{ \rme^{-\frac{2L^2+r_1^2+r_2^2}{2 w^2}} }{8 \pi ^2w^4}\nn 
 &&\times \left[ I_0\left(\frac{L}{w^2}|\r_1+\r_2|  \right)+ I_0\left(\frac{L}{w^2}|\r_1-\r_2|  \right) \right.\nn
&&\left.\qquad\qquad\qquad \qquad\qquad - 2I_0\left( \frac{Lr_1}{w^2} \right)I_0\left( \frac{Lr_2}{w^2} \right)\right].
\eeq
A plot of $G_c(r,r,\varphi_{12})$ is displayed in Fig.~\ref{fig:Gc_dumbbell}, for $r$ close to $L$. The top panel exhibits the angular correlations expected at large deformation ($w/L=0.2$). There is also a large plateau in between the peaks, which gradually softens as the deformation decreases. As we shall argue shortly, the regions of the peaks are dominated by the correlated part, $G(\r_1,\r_2)$, while the plateau corresponds to uncorrelated regions, corresponding to the second term in Eq.~(\ref{eq:Sr1r2alpha12}). The bottom panel shows the results for a smaller deformation parameter ($w/L=0.5$), with the appearance of a $\cos (2\varphi_{12})$ modulation.

The value of the correlation function in the plateau region around $\varphi_{12}=\pi/2$ can be easily estimated. For the particular case $r_1=r_2=r$ and $\cos\varphi_{12}=0$ we have (with here $G_c(r,r,\pi/2)\mapsto G_c(r)$)
\beq\label{eq:Srra12}
G_c(r)=\frac{ \rme^{-\frac{L^2+r^2}{ w^2}} }{ 4\pi ^2w^4}\left\{  I_0\left(\frac{ L r \sqrt{2}}{w^2}  \right) - I_0\left(\frac{Lr}{w^2}\right)\, I_0\left(\frac{Lr}{w^2}\right) \right\}.\nn
\eeq
For large deformation, $w/L\ll 1$, we can use the asymptotic expression of the Bessel function ($I_0(x)\simeq \rme^{x}/\sqrt{2 \pi x}$) to obtain
\beq\label{eq:Srrw02}
G_c(r) &\simeq &-\frac{ \rme^{-\frac{(L-r)^2}{ w^2}} }{ 8\pi ^3 w^2 L r}\left\{ 1-  \frac{\sqrt{2 \pi L r} }{ 2^{1/4\, w}}\rme^{-(2-\sqrt{2}) \frac{Lr}{w^2} } \right\}\nn
&\simeq &-\frac{ \rme^{-\frac{(L-r)^2}{ w^2}} }{ 8\pi ^3 w^2 L r},
\eeq
where, in the second line, we have dropped the subleading term originating from the correlated part of $G_c(r)$. As announced, the plateau regions where $G_c(r)$ takes negative values correspond to uncorrelated regions, where the second contribution to $G_c(r)$ in Eq.~(\ref{eq:Srra12}) dominates.
\begin{figure}
    \centering
    \includegraphics[width=.9\linewidth]{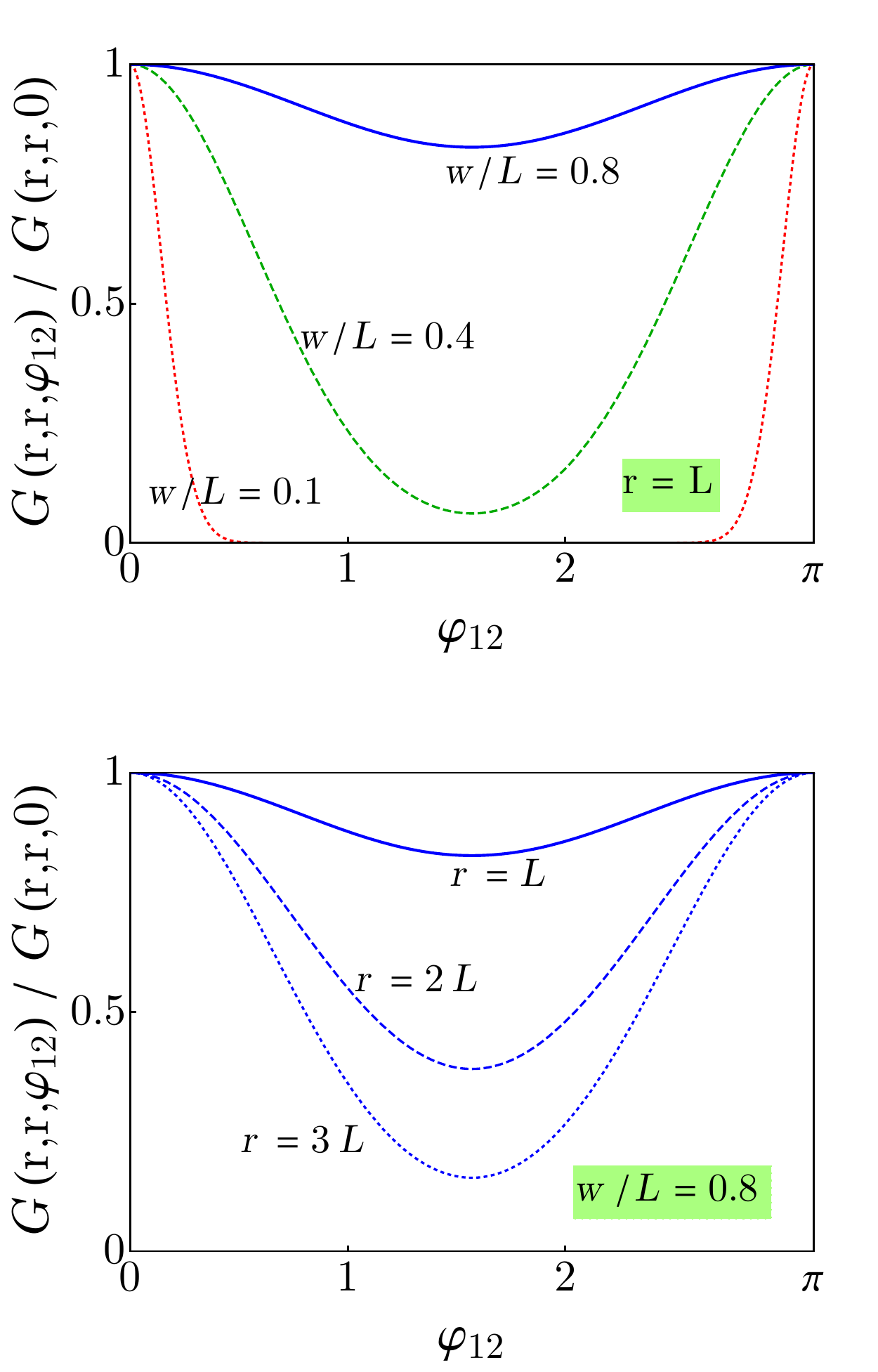}
    \caption{Correlated part of the connected two-point function, $G(r,r,\varphi_{12})$, as a function of $\varphi_{12}$, normalized by its value at $\varphi_{12}=0$. Top: results for $r=L$ and different choices of the deformation parameter, $w=0.8L$ (thick), $w=0.4L$ (dashed), $w=0.1L$ (dotted). Bottom: results for $w=0.8L$ and different choices of $r$, namely, $r=L$ (thick), $r=2L$ (dashed), $r=3L$ (dotted). Note that the uppermost curves (thick lines) are the same in both panels.}
    \label{fig:dumbell4}
\end{figure}

The regions around the peaks, on the other hand, are dominated by the correlated part of (\ref{eq:Srra12}).
Let us then set
\beq
&&G(r,r,\varphi_{12})=\frac{ \rme^{-\frac{L^2+r^2}{ w^2}} }{ 4\pi ^2w^4}\nn && \times\left[ I_0\left(K\sqrt{1-\cos\varphi_{12} } \right)+ I_0\left(K\sqrt{1+\cos\varphi_{12}}  \right)\right],
\eeq
where $K\equiv Lr\sqrt{2}/w^2$.
The peaks occur for $\cos\varphi_{12} =\pm 1$, i.e., for $\varphi_{12}=0$ or $\pi$. Thus, the interplay of the two Bessel functions produces a modulation with a period $\pi$, quite similar to that found for the thin rod (Eq.~(\ref{eq:2ptdumbbell})). The function $G(r,r,\varphi_{12})$ has a minimum at $\varphi_{12}=\pi/2$, where we have  
\beq
G(r,r,\pi/2)=I_0\left(\frac{Lr \sqrt{2}}{w^2}\right)\,\frac{ \rme^{-\frac{L^2+r^2}{ w^2}} }{ 2\pi ^2w^4}.
\eeq
This angular dependence is illustrated in Fig.~\ref{fig:dumbell4} where the quantity ${G(r,r,\varphi_{12})}/{G(r,r,0)}$ is plotted as a function of $\varphi_{12}$ for two values of the deformation, and different choices of $r$.

As the deformation decreases, the angular correlation remains visible but, at fixed $r$, the amplitude of the modulation is much reduced, getting closer to the characteristic $\cos(2\varphi_{12})$ modulation, as can be seen in the upper panel of Fig.~\ref{fig:dumbell4}.  However, as $r$ increases for a fixed moderate deformation, the amplitude of the angular modulation becomes more pronounced, as clearly illustrated in the lower panel of Fig.~\ref{fig:dumbell4}. Intuitively, we understand this feature as a result of the fact that, when the deformation is small, the angular location of points far from the center of the dumbbell becomes better defined at large $r$.

\subsubsection{Fourier transforms}
\begin{figure}[t]
    \centering
    \includegraphics[width=\linewidth]{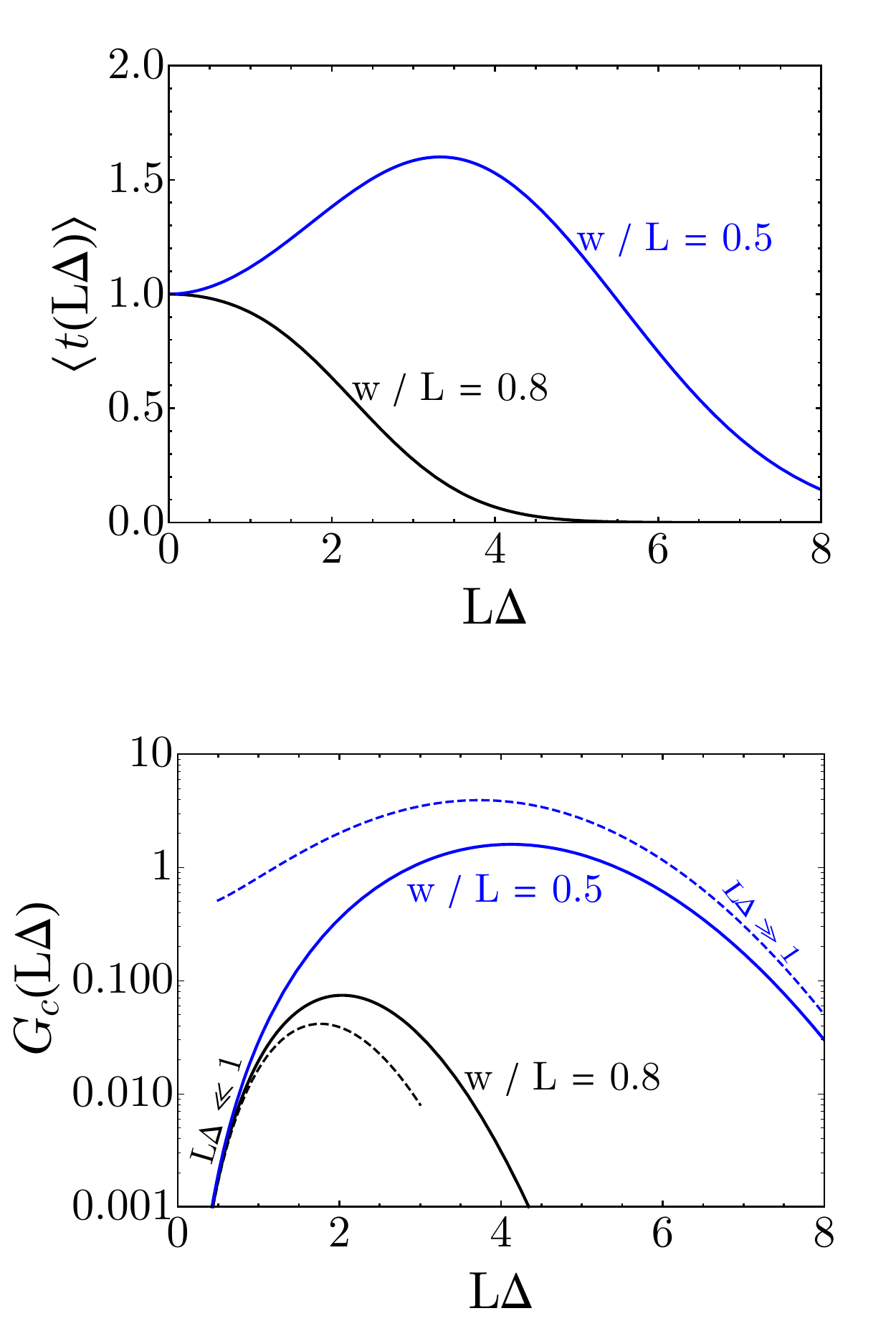}
    \caption{Top: Fourier transform of the one-body density after average over all azimuthal orientations, as a function of $L\Delta $, plotted for a large deformation $w/L=0.5$, and for a small deformation, $w/L=0.8$. Bottom: Fourier transform of the connected two-point function, $G_c(\Delta)$, for the same choices of deformation parameters. The dashed lines indicate the asymptotic limits for $L\Delta \ll 1$ and $L\Delta  \gg 1$, respectively, as discussed in the text.}
    \label{fig:5}
\end{figure}

Let us now see how these properties of both the average thickness function and of the correlation function manifest themselves in their Fourier transforms, which are the quantities that are eventually accessible experimentally (see Sect.~\ref{sec:4}). The Fourier transform of $t(x,y)$, after averaging over the angle $\phi$, is given by  
\beq\label{eq:tildeTADel}
\langle  t(\Delta) \rangle = \rme^{-\frac{1}{2} w^2 \Delta^2} I_0(L\Delta ).
\eeq
For $\Delta=0$, $\langle  t(\Delta) \rangle=1$, which is the value of the integral of the density of the dumbbell. For small $\Delta$, i.e. $L\Delta \ll 1$ and $w \Delta \ll 1$  we have
\beq
\langle  t(\Delta) \rangle\simeq 1 +\frac{1}{2}\left( \frac{1}{2} -\frac{w^2}{L^2} \right) (L\Delta)^2,
\eeq
and we consider two cases of a large and a small deformation parameter. For a large deformation, $w\ll L$, $\langle  t(\Delta) \rangle$ increases with increasing $\Delta$, while the opposite behavior occurs when $w> L/\sqrt{2}$. Thus, the value $w/L=1/\sqrt{2}\approx0.7$ separates the regions of small and large deformations. Now, for $w/L>0.7$, the Gaussian factor $\rme^{-(w^2/2L^2)(L \Delta )^2}$ does not change much until $L\Delta \sim L/w \gg 1$. We may then use the asymptotic value of the Bessel function to get
\beq
\langle  t(\Delta) \rangle\simeq \rme^{-\frac{1}{2} w^2 \Delta^2}\,\rme^{L\Delta }\sqrt{ \frac{1}{2\pi L\Delta }  }.
\eeq
This function has a peak for $L\Delta =L^2/w^2$, i.e., the value of $L\Delta $ at the peak grows with the deformation. We stress that this peak at finite $\Delta$ exists only for large deformation. For small deformation ($w/L<0.7$), the behavior of $\langle t(\Delta) \rangle$ is dominated by the Gaussian factor $\rme^{-(w^2/2L^2)(L \Delta )^2}$ and the peak sits at $\Delta=0$, its width decreasing as the deformation decreases.
These properties are illustrated in the upper panel of Fig.~\ref{fig:5}.

Consider now the Fourier transform of the correlation function. After averaging over the azimuthal orientation, one gets
\beq
&&G_c(\Delta)=\frac{1}{2} \rme^{-w^2 \Delta^2} \left[1+I_0(2L\Delta )-2   I_0(L\Delta )^2 \right].
\eeq
This function is plotted in the lower panel of Fig.~\ref{fig:5}.
The analysis of this expression follows the same lines as that of the thickness function $\langle t(\Delta)\rangle. $ For small deformation corresponding to $L\Delta \ll 1$, we may use $I_0(L\Delta )\simeq 1+(L\Delta )^2/4$ to obtain  
\beq
G_c(\Delta)\simeq \frac{(L\Delta )^4}{32}\,\rme^{-w^2 \Delta^2}.
\eeq
This expression has a maximum for $L\Delta =\sqrt{2}L/w$. 
However, as the deformation increases, the maximum shifts to higher values, and for $L\Delta  \gg 1$, so that a better estimate is given by the asymptotic form of the Bessel functions. One obtains 
\beq\label{eq:SDeltaDB}
G_c(\Delta)&=& \simeq \frac{1}{2} \rme^{-w^2 \Delta^2}\,\rme^{2 L\Delta }\,\left[ \frac{1}{\sqrt{4\pi L \Delta  } } -\frac{2}{2\pi L\Delta  }\right]\nn &\approx& \frac{1}{ \sqrt{16\pi L\Delta }} \, \rme^{-w^2 \Delta^2}\,\rme^{2 L\Delta },
\eeq
where in the last line, we have dropped the contribution of the uncorrelated part of $G_c$, which is subleading. The maximum of $G_c(\Delta)$ now occurs for $L\Delta =L^2/w^2 $. The corresponding value of $G_c$ is 
\beq
G_c(\Delta=L/w^2)=\frac{1}{\sqrt{\pi}}\,\frac{w}{L}\,\rme^{\frac{L^2}{w^2}}.
\eeq
Thus, in all regimes of deformation the peak in $G_c(\Delta)$ informs us on the deformation.

\subsection{Deformed Gaussian}

\begin{figure}[t]
    \centering
    \includegraphics[width=.95\linewidth]{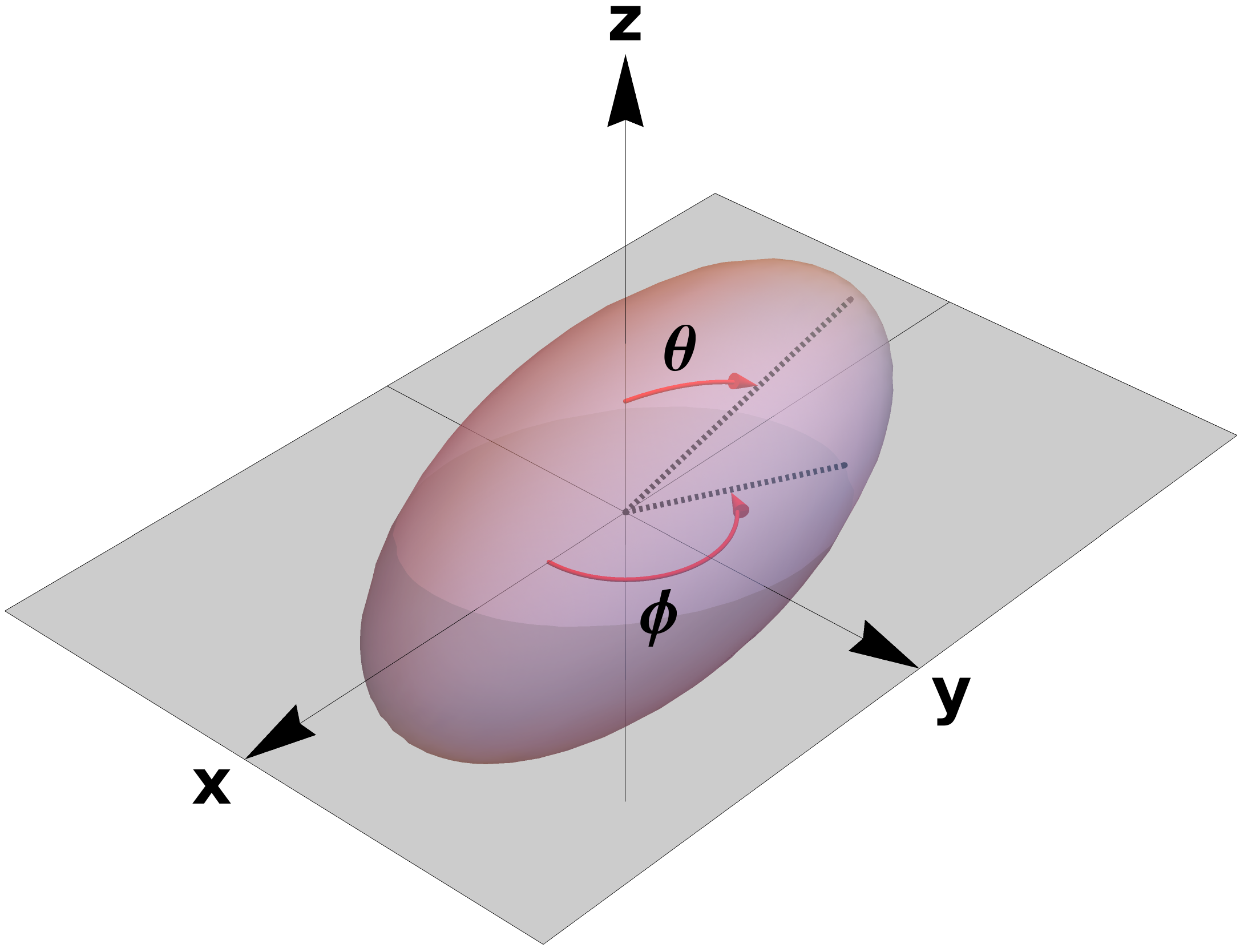}
    \caption{Illustration of an ellipsoidal Gaussian density rotated with respect to the lab frame. The rotation involves a polar tilt, $\theta$, and an azimuthal spin, $\phi$.}
    \label{fig:gauss}
\end{figure}

We now consider the slightly more realistic case of a fictitious deformed nucleus with the following density
\beq
\rho(x,y,z)=\frac{\alpha_z\alpha_x^2}{\pi^{3/2} }\, \rme^{-\alpha_x^2 (x^2+y^2)}\rme^{-\alpha_z^2 z^2}. 
\eeq
This density is normalized to unity and is illustrated in Fig.~\ref{fig:gauss}. It is convenient to characterize the deviation from the spherical shape by a deformation parameter $\delta$, related to $\alpha_x$ and $\alpha_z$ by
\beq
\alpha_x^2=\alpha_0^2 \rme^{\delta/3},\qquad \alpha_z^2=\alpha_0^2\rme^{-2\delta/3}, 
\eeq
so that $\alpha_x^2\alpha_z=\alpha_0^3$ is independent of $\delta$. Note that for $\delta\lesssim 5$ we have, to a good approximation, $\alpha_x^2-\alpha_z^2=\alpha_0^2 \,\delta$. 
In what follows, numerical results are obtained via a phenomenological value of the nuclear radius, equal to that we use for the nucleus $^8$Be in Sec.~\ref{sec:4}, namely
$\alpha_0 = 0.695~{\rm fm}^{-1}.$

\subsubsection{Thickness function}

The thickness function for a given random orientation $\Omega=(\theta,\phi)$ is given by
\beq\label{eq:TAxythetaphiGM}
t_\Omega(x,y)=C_0(\theta) \rme^{B(\theta,\phi, x,y)},
\eeq
where
\beq
C_0(\theta) =\frac{\alpha_x^2 \alpha_z}{\pi ( \alpha_z^2\cos^2\theta+\alpha_x^2\sin^2\theta )^{1/2}},
\eeq
\beq
B&=& -\alpha_x^2\frac{\alpha_z^2 C_1^2 +(\alpha_z^2 \cos^2\theta+\alpha_x^2 \sin^2\theta) C_2^2}{\alpha_z^2 \cos^2\theta+\alpha_x^2 \sin^2\theta}\nn
&=&-\alpha_x^2 C_2^2-\frac{\alpha_x^2 \alpha_z^2 }{\alpha_z^2 \cos^2\theta+\alpha_x^2 \sin^2\theta}C_1^2 \nn
&=&b_1 \,C_1^2+b_2\, C_2^2, 
\eeq
and 
\beq
C_1\equiv x\cos\phi +y\sin\phi,  \quad C_2\equiv y\cos\phi - x \sin\phi.
\eeq
The average over the random orientation of the nucleus is made in two steps. One first average over the angle $\phi$, keeping $\theta$ fixed. One then obtains an expression similar to (\ref{eq:thickdumb}), namely
\beq\label{eq:thickaverage}
\langle t_\theta(r)\rangle&=&C_0\int\frac{\rmd \phi}{2\pi} \, \rme^{b_1 r^2 \cos^2\phi+b_2 r^2 \sin^2\phi}\nn
&=& C_0\,\rme^{\frac{(b_1+b_2) r^2}{2}} I_0\left( (b_1-b_2){r^2}/{2}  \right) ,
\eeq
with
\beq\label{eq:b1andb2}
&& b_1-b_2=\frac{\alpha_x^2(\alpha_x^2-\alpha_z^2) \sin^2\theta }{\alpha_z^2\cos^2\!\theta+\alpha_x^2\sin^2\!\theta}\nn
&&\frac{b_1+b_2}{2}=-\frac{ \alpha_x^2}{2}\left[\frac{2 \alpha_z^2 +(\alpha_x^2-\alpha_z^2)\sin^2\theta}{ \alpha_z^2\cos^2\!\theta+\alpha_x^2\sin^2\!\theta }\right].
\eeq
The second step is the average over the angle $\theta$. This latter step does not seem to be doable analytically, but the numerical evaluation is straightforward.

\subsubsection{Density-density correlation function }
\begin{figure}[t]
\begin{center}
\includegraphics[width=0.95\linewidth]{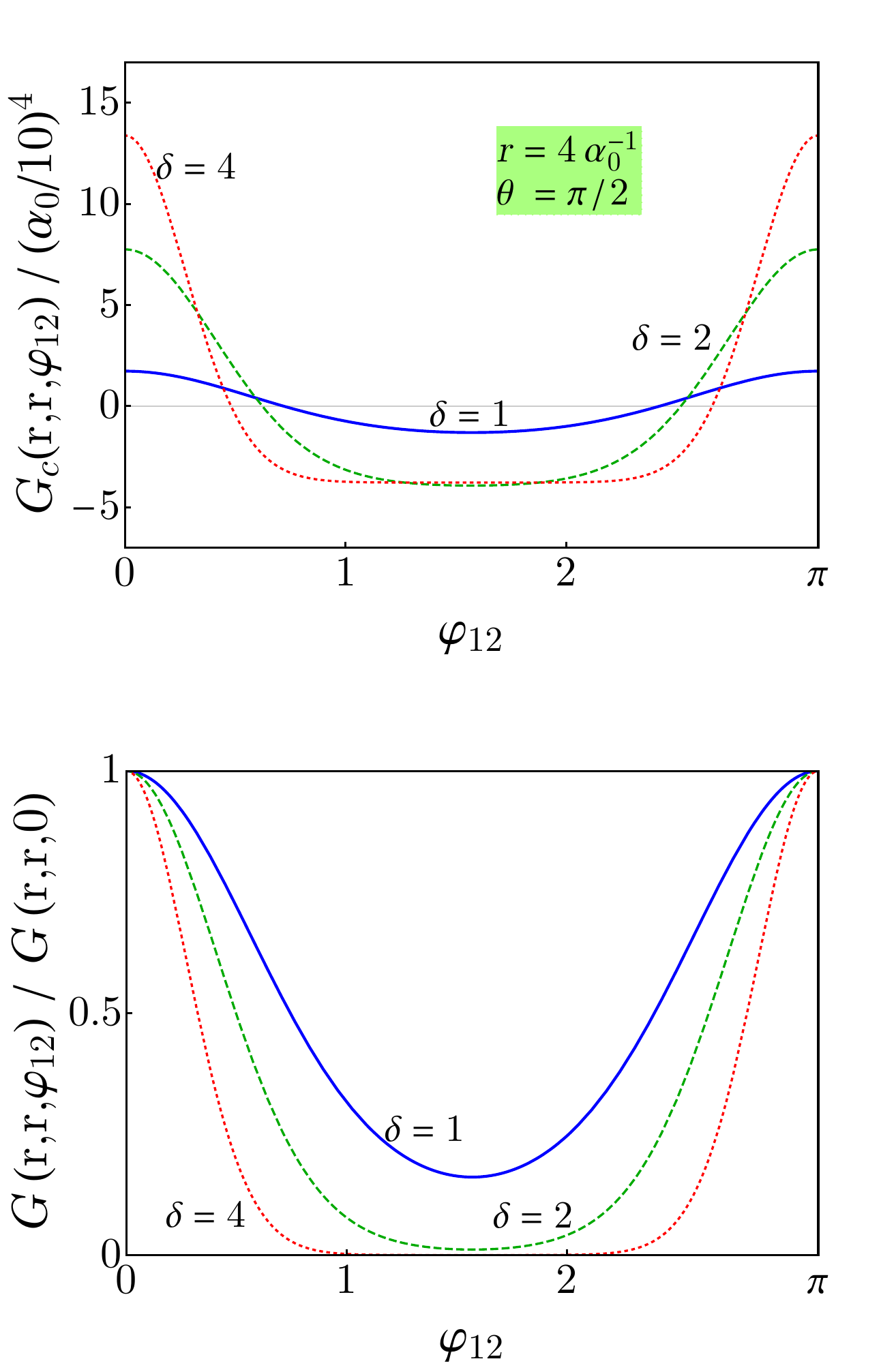}
\caption{Top: Dimensionless connected two-point function $G_c(r,r,\varphi_{12})$ as a function of $\varphi_{12}$ for various values of $\delta$, namely, $\delta=1$ (thick), $\delta=2$ (dashed), $\delta=4$ (dotted). Note that, for $\delta=0$, the root mean squared radius of the Gaussian density is $\alpha_0^{-1}$. Here we set $r=4/\alpha_0$. Bottom: Same plot for the correlated part $G(r,r,\varphi_{12})$ rescaled by its value at $\varphi_{12}=0$. }
\label{fig:7}
\end{center}
\end{figure}

The calculation of the density-density correlation function proceeds as for the dumbbell, and yields similar expressions. After averaging over $\phi$, one gets
\beq
G_c(r_1,r_2)=C_0^2\, \rme^{(b_1+b_2)\frac{r_1^2+r_2^2}{2}} h(r_1,r_2,\varphi_{12}),
\eeq
where
\beq
&&h(r_1,r_2,\varphi_{12})= I_0\left( (b_1-b_2)\sqrt{ f(\varphi_{12}, r_1,r_2)} \right)\nn &&- I_0\left( (b_1-b_2)r_1^2/2 \right)  \, I_0\left( (b_1-b_2)r_2^2/2 \right) ,
\eeq
and 
\beq
f(r_1,r_2,\varphi_{12})=\frac{r_1^4+r_2^4}{4}+\frac{r_1^2 r_2^2}{2} \cos 2\varphi_{12} . 
\eeq
Setting $\theta=\pi/2$, we plot the function $G_c(r,r,\varphi_{12})$, rescaled by $\alpha_0^4$, in the upper panel of Fig.~\ref{fig:7}. We see the characteristic peaks around $\varphi_{12}=0$ and $\pi$ for a large deformation ($\delta=4$), which become smeared into the characteristic $\cos(2\varphi_{12})$ modulation for a smaller deformation parameter ($\delta=1$), as already observed for the dumbbell. The lower panel of Fig.~\ref{fig:7} shows instead the trend of the correlated part only, $G(r,r,\varphi_{12})$, akin to the plot shown in Fig.~\ref{fig:dumbell4}.  

\begin{figure}[t]
\begin{center}
\includegraphics[width=.95\linewidth]{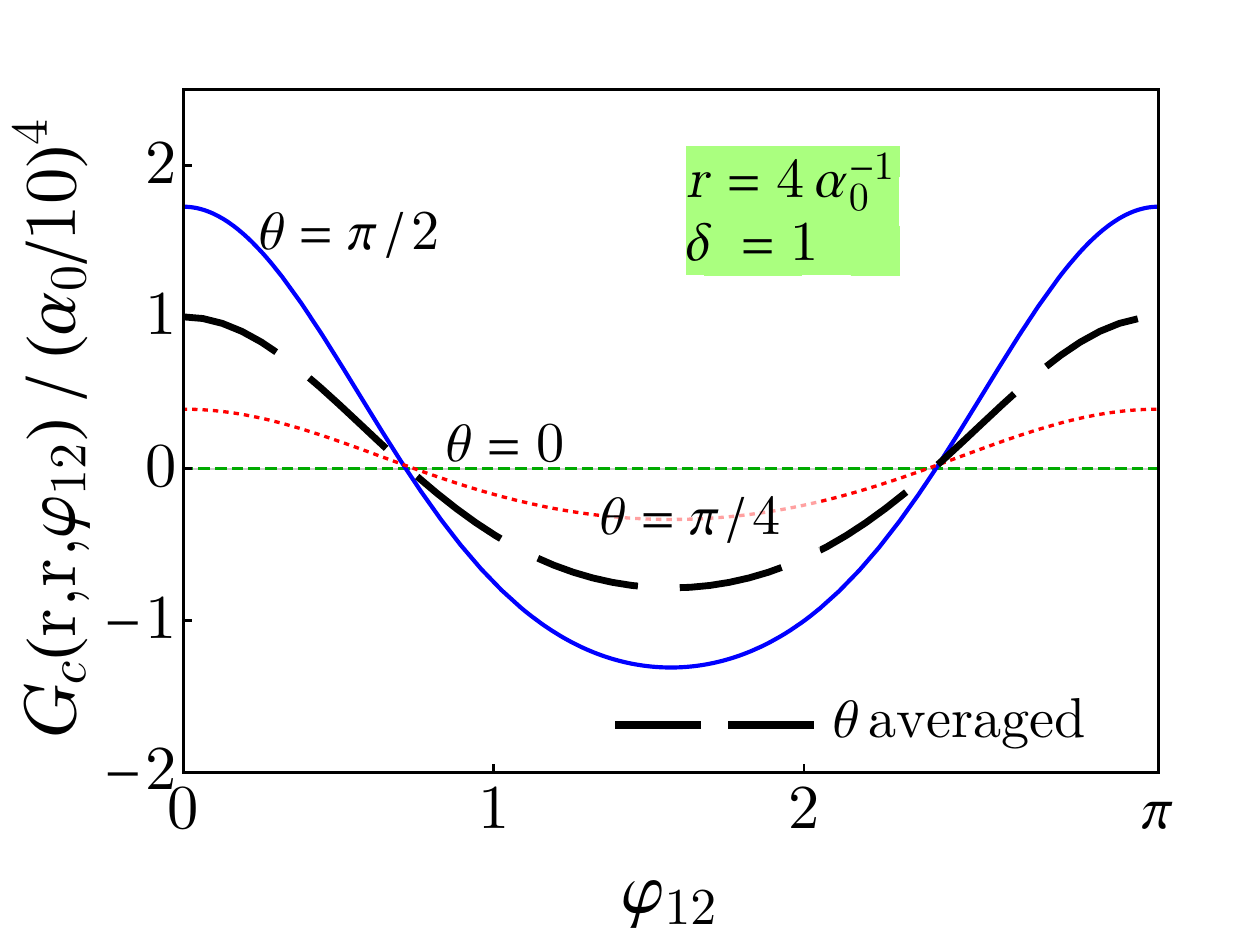}
\caption{The correlation function $G_c(r,r,\varphi_{12})$ rescaled by $\alpha_0^4$ as a function of $\varphi_{12}$ for $\delta=1$, $r=4/\alpha_0$, and various values of $\theta$. Note that the curve for $\theta=\pi/2$ corresponds to the thick lines in Fig.~\ref{fig:7}. The black dashed line represents the average over $\theta$. }
\label{fig:Gaussian_Stheta}
\end{center}
\end{figure}

The regime of small deformations is the most relevant for realistic applications. In this regime, one  obtains a simple relation for $h(r_1,r_2,\varphi_{12})$. 
We first note that $b_1-b_2$ is proportional to the deformation ($\alpha_x^2-\alpha_z^2\simeq\alpha_0^2 \,\delta$), that is,
\beq
 b_1-b_2\simeq \alpha_0^2 \sin^2\!\theta\, \delta.
 \eeq
Thus, using $I_0(x)\simeq 1+x^2/4$, we get
\beq \label{eq:hsmalldef}
h(r_1,r_2,\varphi_{12})\simeq\frac{\delta^2 \sin^4\!\theta}{8} \,\alpha_0^4r_1^2 r_2^2\,  \cos 2\varphi_{12} \,.
\eeq
As expected, this vanishes when $\theta=0$ regardless of the value of $\delta$ since, for $\theta=0$, the thickness function is symmetric in the transverse plane $(x,y)$. The dependence of $h(r_1,r_2,\varphi_{12})$ on $\varphi_{12}$  explicitly exhibits the modulation in $\cos 2\varphi_{12}$. Furthermore, Eq.~(\ref{eq:hsmalldef})  shows that the amplitude of such a modulation increases for higher values of the radial coordinates. In addition, Fig.~\ref{fig:Gaussian_Stheta} illustrates the role of $\theta$ in controlling the apparent deformation of the thickness function and how this affects $G_c(r,r,\varphi_{12})$: As $\theta\to 0$, the thickness function becomes invariant under rotations in the plane $(x,y)$ and the angular correlation disappears, while for $\theta=\pi/2$ the apparent deformation is maximal and so is the modulation amplitude. 

\begin{figure}[t]
    \centering
    \includegraphics[width=0.95\linewidth]{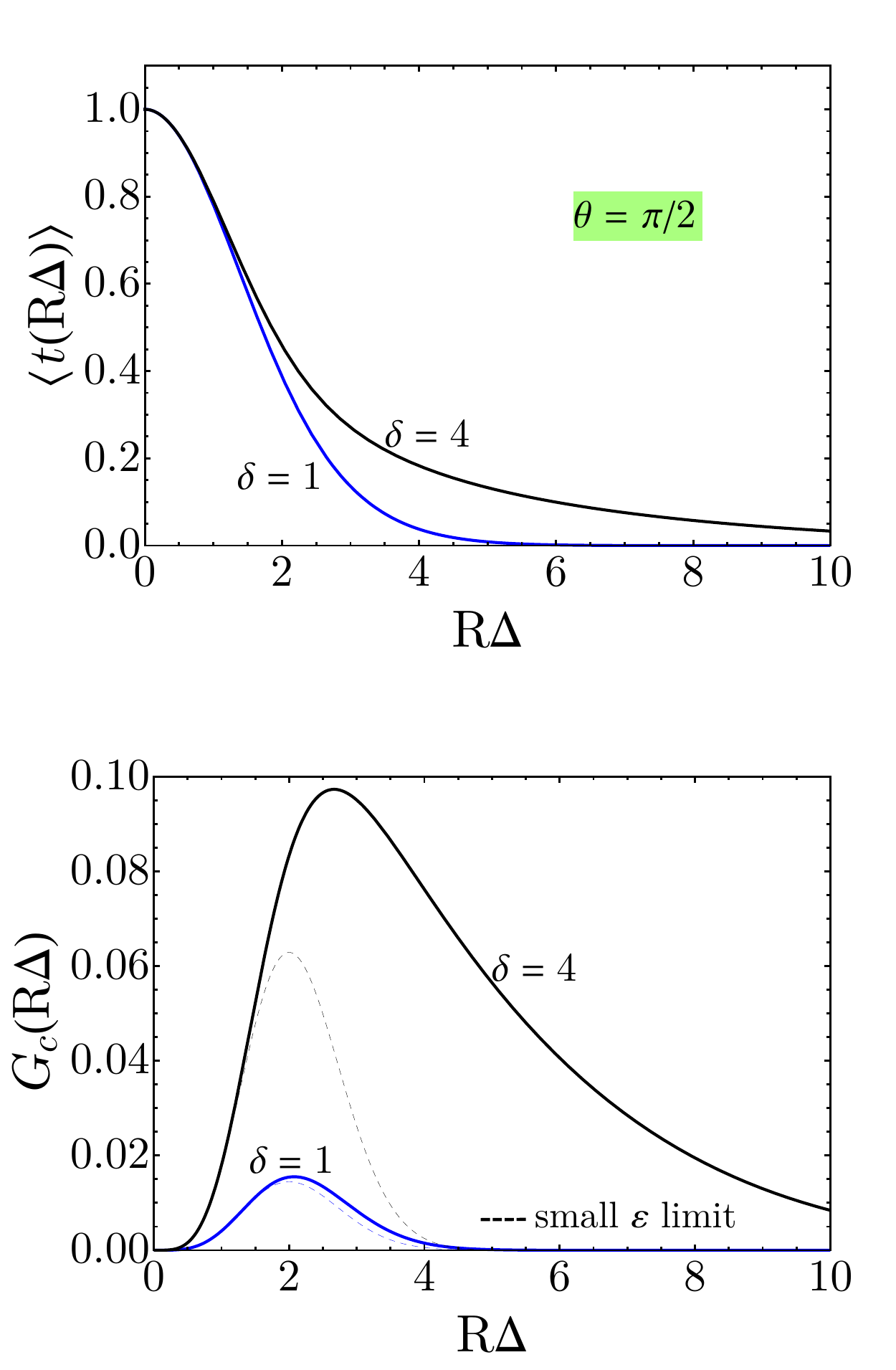}
    \caption{Top: one-body density as a function of $R\Delta $ obtained by averaging the Fourier transform of the Gaussian density over azimuthal orientations for $\theta=\pi/2$. Note that $\delta=1$ corresponds to $\varepsilon=0.46$, while $\delta=4$ corresponds to $\varepsilon=0.96$. Bottom: Value of the connected two-point function in Fourier space $G_c(R\Delta )$ as a function of $R\Delta $. The dashed curves represent the limit behavior for small  $\varepsilon$ described by Eq.~(\ref{eq:Gc_smallEps}).}
    \label{fig:gauss_DR}
\end{figure}

\subsubsection{Fourier transforms}

We now turn to the Fourier transform. For simplicity, we present results only for $\theta=\pi/2$. For the one-point function, we obtain 
\beq
\langle t(\Delta) \rangle=A \,\rme^{-\frac{R^2 \Delta^2}{4}  }I_0\left( \varepsilon \frac{R^2\Delta^2 }{4 } \right) .
\eeq
where 
\beq
 R^2=\langle x^2+y^2\rangle= \frac{e^{-\delta/3}}{2 \,\alpha_0^2}\left ( e^\delta +1 \right)  ,
\eeq
is the mean squared radius of the thickness function $t(x,y)$,  and 
   \beq
   \varepsilon=\frac{\langle x^2-y^2\rangle}{\langle x^2+y^2\rangle}=\frac{ \rme^{ \delta/3}-\rme^{-2\delta/3}}{\rme^{ \delta/3}+\rme^{-2\delta/3}}=\tanh(\delta/2).
   \eeq
   the associated eccentricity. 
The function $\langle t(\Delta) \rangle$ has a maximum at $\Delta=0$, and vanishes at large values of $\Delta$. When plotted as a function of $R \Delta $, it depends only weakly on the deformation, that is, most of the dependence on the value of $\delta$ is carried by $R$ itself, according to the equation above.
Similarly, the correlation function is given by 
\beq
G_c(\Delta)=\rme^{-\frac{R^2\Delta^2}{2} } \! \left[ I_0\left( \varepsilon \frac{R^2\Delta^2}{2}  \right)  -I_0\left( \varepsilon \frac{ R^2\Delta^2}{4 }\right)^2\right].\nn
\eeq
The one-point function, $\langle  t (\Delta) \rangle$ and the two-point function $G_c(\Delta)$ are shown in Fig.~\ref{fig:gauss_DR} as a function of $R \Delta $ for two choices of $\delta$.

To capture the behavior in the vicinity of the peak of $G_c(\Delta)$, for small deformations, it is sufficient to expand the Bessel functions to quadratic order.  We get  then
\beq
I_0\left(  \frac{Q}{2}\Delta^2  \right)  -I_0\left(  \frac{ Q}{4 }\Delta^2\right)^2 \simeq \varepsilon^2 \frac{R^4\Delta^4}{32}.
\eeq
This implies that 
\beq
\label{eq:Gc_smallEps}
G_c(\Delta)\simeq \frac{\varepsilon^2}{8} \left(\frac{R^2\Delta^2}{2}\right)^2\rme^{-\frac{R^2}{2} \Delta^2}.
\eeq
This function has a maximum at $R\Delta =2$, and is valid for small values of $\varepsilon$ (typically $\varepsilon\lesssim 0.5$). The value of $G_c$ at the maximum is then proportional to the deformation of the system 
\beq
G_c^{\text max}=\frac{\varepsilon^2}{2 \rme^2}.
\eeq
In Fig.~\ref{fig:gauss_DR}, these limits are shown as dashed lines. We see that Eq.~(\ref{eq:Gc_smallEps}) captures accurately the curve for $\delta=1$ ($\varepsilon=0.46$). For higher deformations, the Bessel functions significantly modify the behavior of $G_c(\Delta)$ for larger values of $\Delta$.

%%%%%%%%%%%%%%%%%%%%%%%%%%%%%%%%%%%%%%%%%%%%%%%%%%%%%%
%%%%%%%%%%%%%%%%%%%%%%%%%%%%%%%%%%%%%%%%%%%%%%%%%%%%%%
%%%%%%%%%%%%%%%%%%%%%%%%%%%%%%%%%%%%%%%%%%%%%%%%%%%%%%
%%%%%%%%%%%%%%%%%%%%%%%%%%%%%%%%%%%%%%%%%%%%%%%%%%%%%%
%%%%%%%%%%%%%%%%%%%%%%%%%%%%%%%%%%%%%%%%%%%%%%%%%%%%%%
%%%%%%%%%%%%%%%%%%%%%%%%%%%%%%%%%%%%%%%%%%%%%%%%%%%%%%
%%%%%%%%%%%%%%%%%%%%%%%%%%%%%%%%%%%%%%%%%%%%%%%%%%%%%%
%%%%%%%%%%%%%%%%%%%%%%%%%%%%%%%%%%%%%%%%%%%%%%%%%%%%%%
%%%%%%%%%%%%%%%%%%%%%%%%%%%%%%%%%%%%%%%%%%%%%%%%%%%%%%
%%%%%%%%%%%%%%%%%%%%%%%%%%%%%%%%%%%%%%%%%%%%%%%%%%%%%%
%%%%%%%%%%%%%%%%%%%%%%%%%%%%%%%%%%%%%%%%%%%%%%%%%%%%%%
%%%%%%%%%%%%%%%%%%%%%%%%%%%%%%%%%%%%%%%%%%%%%%%%%%%%%%

\section{Diffractive vector meson production as a probe of correlations in Beryllium-8}
\label{sec:4}
 Our last example deals with a physical process, that of diffractive photo-production (or electro-production) of vector mesons on a light nucleus. The calculation to be presented follows the lines initiated by Caldwell and Kowalski \cite{Caldwell:2010zza}, and more recently investigated in a detailed numerical simulation by M\"antysaari \textit{et al.} \cite{Mantysaari:2023qsq}. The example to be discussed remains somewhat academic since it involves the nucleus $^8$Be, which is not a stable nucleus (it decays into two alpha particles \cite{wikipediaBeryllium8}). However, our goal here is not to get realistic estimates of cross sections, but rather to illustrate the effect of angular correlations in a calculation as close as possible to what a realistic calculation would be. In that perspective, $^8$Be is the simplest deformed nucleus. More realistic examples involving light nuclei will be discussed in future works.
 
\subsection{Harmonic oscillator model of light nuclei}

The 3-dimensional harmonic oscillator provides a good approximation for the self-consistent mean field of light nuclei. It also provides a simple illustration of the basic mechanism responsible for the emergence of nuclear deformation in a mean field picture. We use it here to obtain the intrinsic state of $^8$Be.

The Hamiltonian of a particle with mass $m$ confined in a three-dimensional harmonic oscillator reads
\beq
H=\frac{\p^2}{2m} +\frac{m}{2}(\omega_x^2 x^2+\omega_y^2 y^2+\omega_z^2 z^2).
\eeq
The frequencies $\omega_x$, $\omega_y$, $\omega_z$ are treated here as free parameters to be determined via a self-consistency argument. The single particle states are characterized by the three (integer) quantum numbers $n_x,n_y,n_z$, which define their energy
\beq 
\varepsilon_{n_x,n_y,n_z}= \sum_{i=x,y,z} \left( n_i+\frac{1}{2} \right) \hbar \omega_i.
\eeq
We denote the corresponding orbital by $(n_xn_yn_z)$.
The (intrinsic) ground state is obtained by filling the single particle levels with 4 nucleons (two neutrons and two protons) in each level. It is convenient to introduce the total effective number of quanta in each direction. Thus, we introduce
\beq
N_x=\sum_{i=1}^A\left( n_x^{(i)} +\frac{1}{2} \right),
\eeq
where $\sum_{i=1}^A$ runs over the $A$ nucleons. We define similarly $N_y$ and $N_z$. In terms of these numbers, we have
\beq
\langle x^2\rangle= \sum_{i=1}^A \langle x_i^2\rangle=4\,\frac{N_x}{\alpha_x^2}, 
\eeq
where 
\beq
\alpha_x\equiv \sqrt{\frac{m\omega_x}{\hbar}}. 
\eeq
 Similarly we have
\beq
\frac{\langle p_x^2\rangle}{\hbar^2}=4N_x \alpha_x^2,
\eeq
and analogously for $\langle p_{y,z}^2 \rangle$. Given the filling of the orbitals, we can calculate $N_x,N_y,N_z$. The frequencies $\omega_x, \omega_y, \omega_z$ are then determined by the two conditions \cite{bohr}
\beq\label{eq:volumecons}
\omega_x\omega_y\omega_z=\omega_0^3, 
\eeq
and 
\beq\label{eq:equipartition}
N_x\omega_x=N_y\omega_y=N_z\omega_z.
\eeq
The first one expresses the conservation of the volume when the nucleus gets deformed. The second expresses the isotropy of the velocity distribution in the equilibrium state. Note that the volume conservation entails $\alpha_z\alpha_x^2=\alpha_0^3$.
\begin{figure}
    \centering
    \includegraphics[width=.95\linewidth]{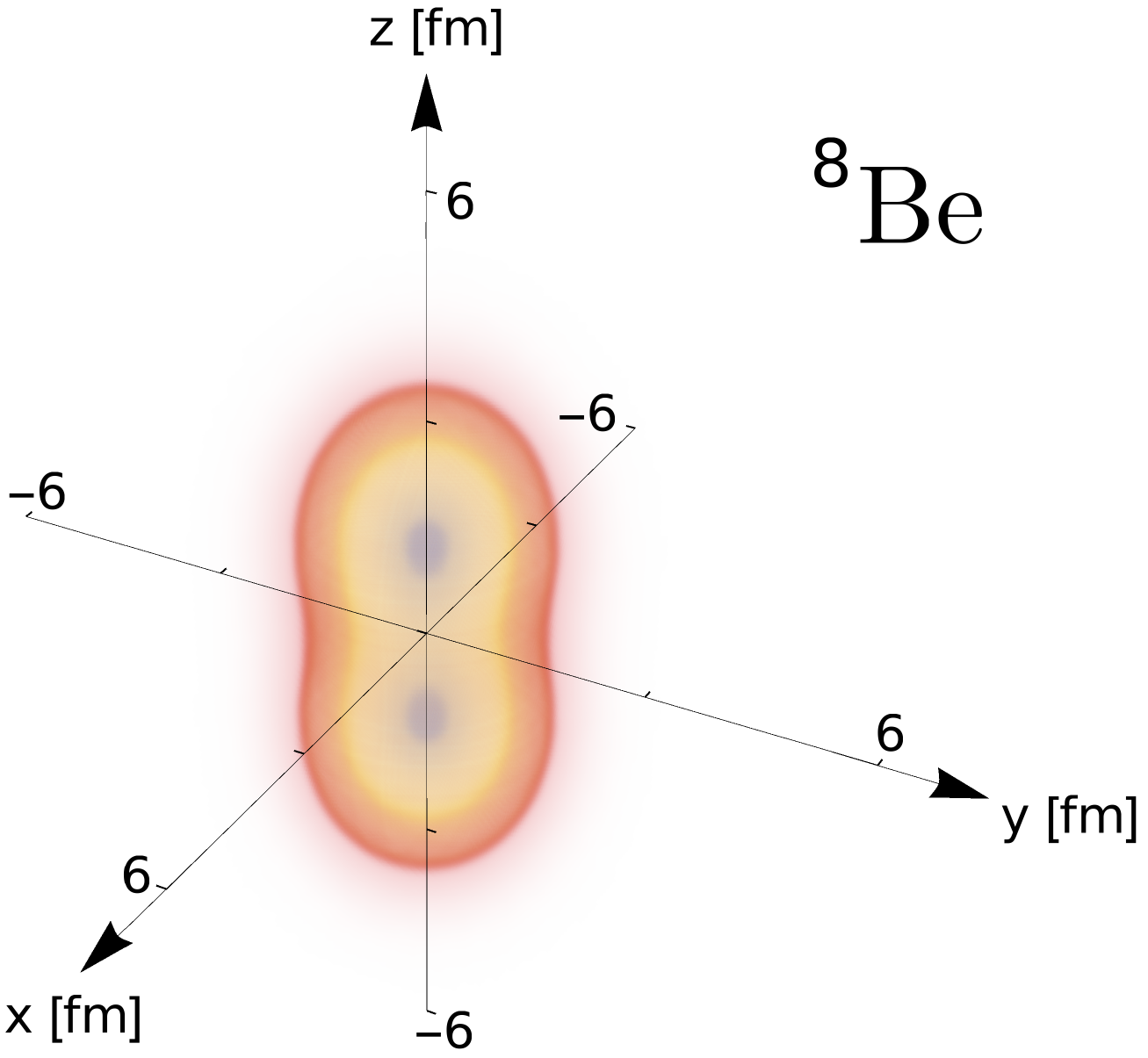}
    \caption{Intrinsic matter density of $^8$Be, as given by Eq.~(\ref{eq:8Bedensity}). The density is higher at the center of the two clusters (blue spots).}
    \label{fig:8Bedens}
\end{figure}

In the case of $^{8}$Be, the two orbitals $(000)$ and $(001)$ are occupied by four nucleons each. We have then\footnote{Note that the parameter $\delta$ is related, for small deformation, to the Bohr quadrupole deformation parameter, $\beta$, by $\delta\simeq 3\beta\sqrt{5/16\pi}$.}
 \beq
 N_x=N_y=1,\; N_z=2,\quad \frac{\omega_x}{\omega_z}=2=\rme^\delta,\; \delta=0.69.
 \eeq 
 The asymmetric filling of the first excited level of the oscillator (out of the three degenerate orbitals (100), (010) and (001), only the last one is occupied) is at the origin of the deformation of the intrinsic state and of the clustering of the density into what resembles two alpha particles, which is clearly visible in Fig.~\ref{fig:8Bedens}. This clustering is therefore not due here to specific interactions which would favor the formation of alpha particles within the nucleus.  

The (normalized) wave functions that we need are those corresponding to the ground state ($\varphi_0$) and the first excited state ($\varphi_1$) of the one-dimensional oscillator. These are given respectively by
\beq
&&\varphi_0(x)=\frac{\alpha^{1/2}}{\pi^{1/4}}{\rm e}^{-\alpha^2 x^2/2}, \nn
&&\varphi_1(x)=\frac{\alpha^{1/2}}{(4\pi)^{1/4}}{\rm e}^{-\alpha^2 x^2/2} \,2\alpha x .
\eeq
The density of the nucleus in the intrinsic frame is then given by $\rho_{\rm Be}=A \rho$, with 
\beq
\label{eq:8Bedensity}
\rho(x,y,z)=\frac{\alpha_z\alpha_x^2}{2\pi^{3/2} }\, \rme^{-\alpha_x^2 (x^2+y^2)}\rme^{-\alpha_z^2 z^2}\left( 1+2\alpha_z^2 z^2 \right),\nn
\eeq
such that $\int \rmd x \rmd y \rmd z\,\rho(x,y,z)=1$. An illustration of this density is provided in Fig.~\ref{fig:8Bedens}.  To fix the parameters of the model, we take the commonly used phenomenological value
\beq\label{eq:omega0A}
\hbar\omega_0=40 \,A^{-1/3}\:{\rm MeV}.
\eeq
The values of $\alpha_i$ are then (in fm$^{-1}$)
\beq
\alpha_0=0.695,\quad \alpha_x= 0.780,\quad \alpha_z=0.552.
\eeq
With these parameters the root mean squared radius of $^8$Be is $R_{\rm Be} = 2.22~{\rm fm}$.

\subsubsection{Thickness function}

The thickness function is obtained from the integral of $\rho_\Omega(x,y,z)$ over $z$. The calculation proceeds as in the examples of the previous section. It leads to an expression quite similar to that of the Gaussian model, Eq.~(\ref{eq:thickaverage}).
After averaging over $\phi$ we get
\beq
\label{eq:Tr8Be}
&& \langle t_\theta(r) \rangle=\frac{C_0}{16} \rme^{\frac{b_1+b_2}{2}r^2}  \left[  (2 a_0+a_1 r^2) I_0((b_1-b_2)r^2/2)\right. \nn && \left. +a_1 r^2 I_1((b_1-b_2)r^2/2)   \right],
\eeq
where $b_1$ and $b_2$ are given in Eq.~(\ref{eq:b1andb2}) and we have here
\beq
&&C_0 a_0=\frac{4\alpha_x^2\alpha_z}{\pi}\,\frac{2 \alpha_z^2 \cos^2\theta +\alpha_x^2\sin^2\theta}{ (\alpha_z^2\cos^2\!\theta+\alpha_x^2\sin^2\!\theta)^{3/2} },\nn
&& \frac{C_0}{2} a_1=\frac{4 \alpha_z^3\alpha_x^6 \sin^2\theta}{ \pi (\alpha_z^2\cos^2\!\theta+\alpha_x^2\sin^2\!\theta)^{5/2} }.
\eeq
The expression of the thickness function simplifies considerably  for $\theta=\pi/2$, in which case we have
\beq
C_0=\frac{4\alpha_z}{\pi \alpha_x^3},\quad a_0=\alpha_x^4,\quad a_1=2\alpha_x^4 \alpha_z^2, 
\eeq
and
\beq\label{eq:tpi2r}
&&\langle t_{\pi/2}(r) \rangle=\frac{\alpha_x\alpha_z}{2\pi} \rme^{-(\alpha_x^2+\alpha_y^2)r^2/2}  \nn && \left[  (1\!+\!\alpha_z^2 r^2) I_0((\alpha_x^2\!-\!\alpha_y^2)r^2 /2) +\alpha_z^2 r^2 I_1((\alpha_x^2\!-\!\alpha_y^2)r^2/2)   \right].\nn
\eeq

By taking the Fourier transform (for a given $\theta$ and  averaging over $\phi$) one obtains
\beq\label{eq:DeltathetaBe}
\langle t_\theta(\Delta)\rangle=\frac{\rme^{ \frac{\Delta^2}{8 b_+}} }{8}\left[\!
\left( 8\!-\! \frac{\Delta_\theta^2}{\alpha_z^2} \right)I_0\!\left( \frac{\Delta^2}{8 b_-}  \right)\!-\!\frac{\Delta_\theta^2}{\alpha_z^2}I_1\!\left( \frac{\Delta^2}{8 b_-}\right) \!\right], \nn
\eeq
where we have set
\beq
\Delta_\theta^2\equiv\Delta^2\sin^2\theta,\quad \frac{1}{b_+}\equiv\frac{1}{b_1}+\frac{1}{b_2},\quad \frac{1}{b_-}\equiv\frac{1}{b_1}-\frac{1}{b_2}.
\eeq
This quantity is plotted in the upper panel of Fig.~\ref{fig:FT8Be} below. Note that a ``diffraction minimum'' (i.e. a zero of $\langle t_\theta(\Delta)\rangle$) exists only for $\theta\gtrsim\pi/4$.

\subsubsection{Density-density correlation function}

\begin{figure}[t]
\begin{center}
\includegraphics[width=.92\linewidth]{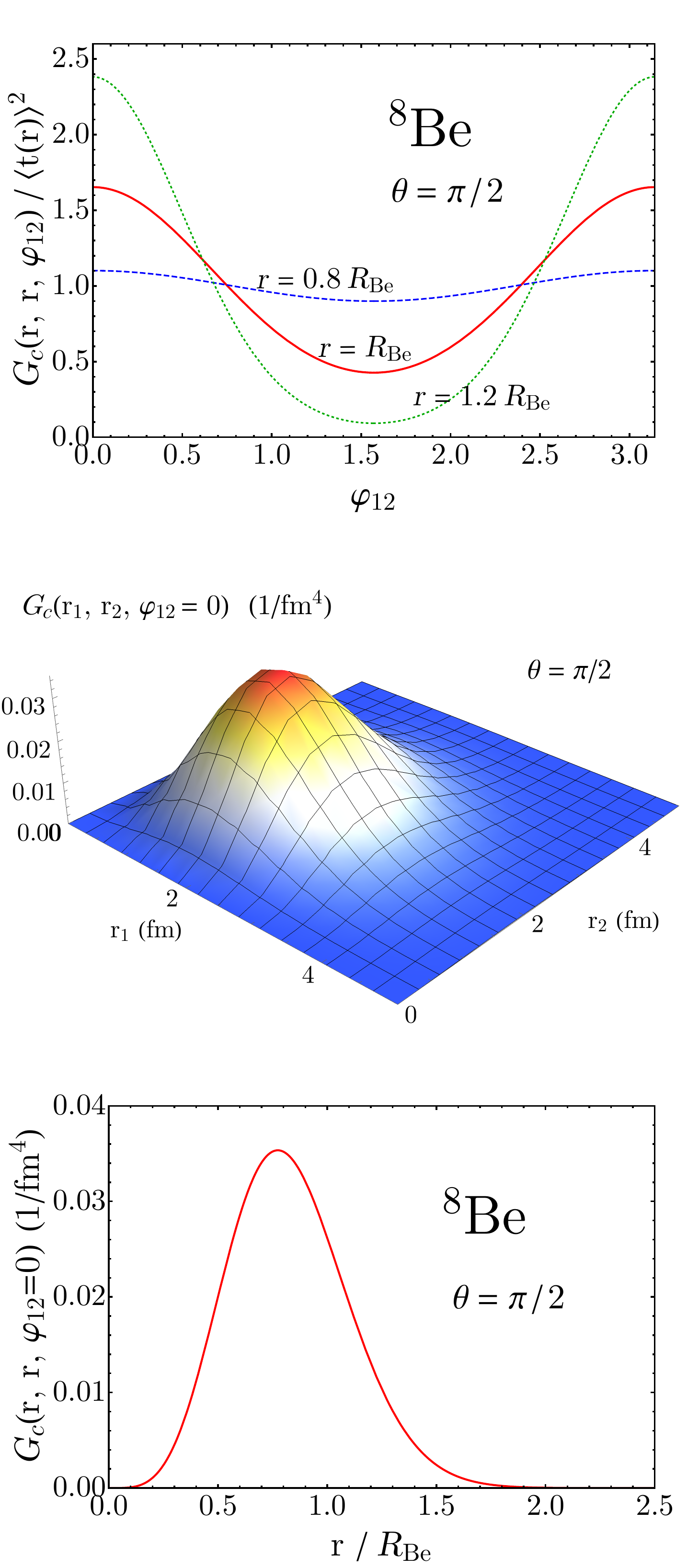}
\caption{We show the connected two-body correlation function $G_c(r_1,r_2,\varphi_{12})$ for the model of $^{8}$Be discussed in the text. Top: plot of $G_c(r,r,\varphi_{12})/\langle t(r)\rangle^2$, where $\langle t(r) \rangle$ corresponds to Eq.~(\ref{eq:tpi2r}). Middle: Plot of $G_c(r_1,r_2,\varphi_{12}=0)$ as a function of both radial coordinates. Lower:  Plot of $G_c(r,r,\varphi_{12}=0)$ as a function of $r/R_{\rm Be}$. The value of the rms radius of the nucleus is $R_{\rm Be}=2.22$ fm.}
\label{fig:8BeGG}
\end{center}
\end{figure}

The calculations of the coordinate space correlation functions $G(r_1,r_2,\varphi_{12})$ and $G_c(r_1,r_2,\varphi_{12})$ lead to complicated expressions which we do not present here. However, these expressions are easy to evaluate numerically, and the salient properties of the resulting correlation functions are illustrated in Fig.~\ref{fig:8BeGG}. In the upper panel, the function $G_c(r,r,\varphi_{12})$ divided by $\langle t(r)\rangle^2$, is plotted for $\theta=\pi/2$, as a function of $\varphi_{12}$. We recognize the by-now familiar modulation in $\cos (2\varphi_{12})$, whose amplitude grows as the value of $r$ increases. The comparison to the models of the previous section suggests that $^{8}$Be behaves as a  mildly-deformed system (even though, when compared to other deformed nuclei, the corresponding deformation parameter would be considered as large). Therefore, we expect such an azimuthal modulation of the correlation function to be a characteristic property of the ground state of any axially symmetric nucleus exhibiting rotational behavior.

Further aspects of the correlations can be seen in the middle and lower panels of Fig.~\ref{fig:8BeGG}. The middle panel displays the correlation function in the vicinity of the angular peak, namely $G_c(r_1,r_2,\varphi_{12}=0)$. We see that the angular correlation is maximal when both $r_1$ and $r_2$ are of the order of $r=R_{\rm Be}$, reflecting the long range nature of the angular correlations. This is confirmed in the lower panel that displays $G_c(r,r,\varphi_{12}=0)$ as a function of $r/ R_{\text{Be}}$. This property also manifests itself in the Fourier transform, to which we now turn.

\begin{figure}[t]
\begin{center}
\includegraphics[width=0.9\linewidth]{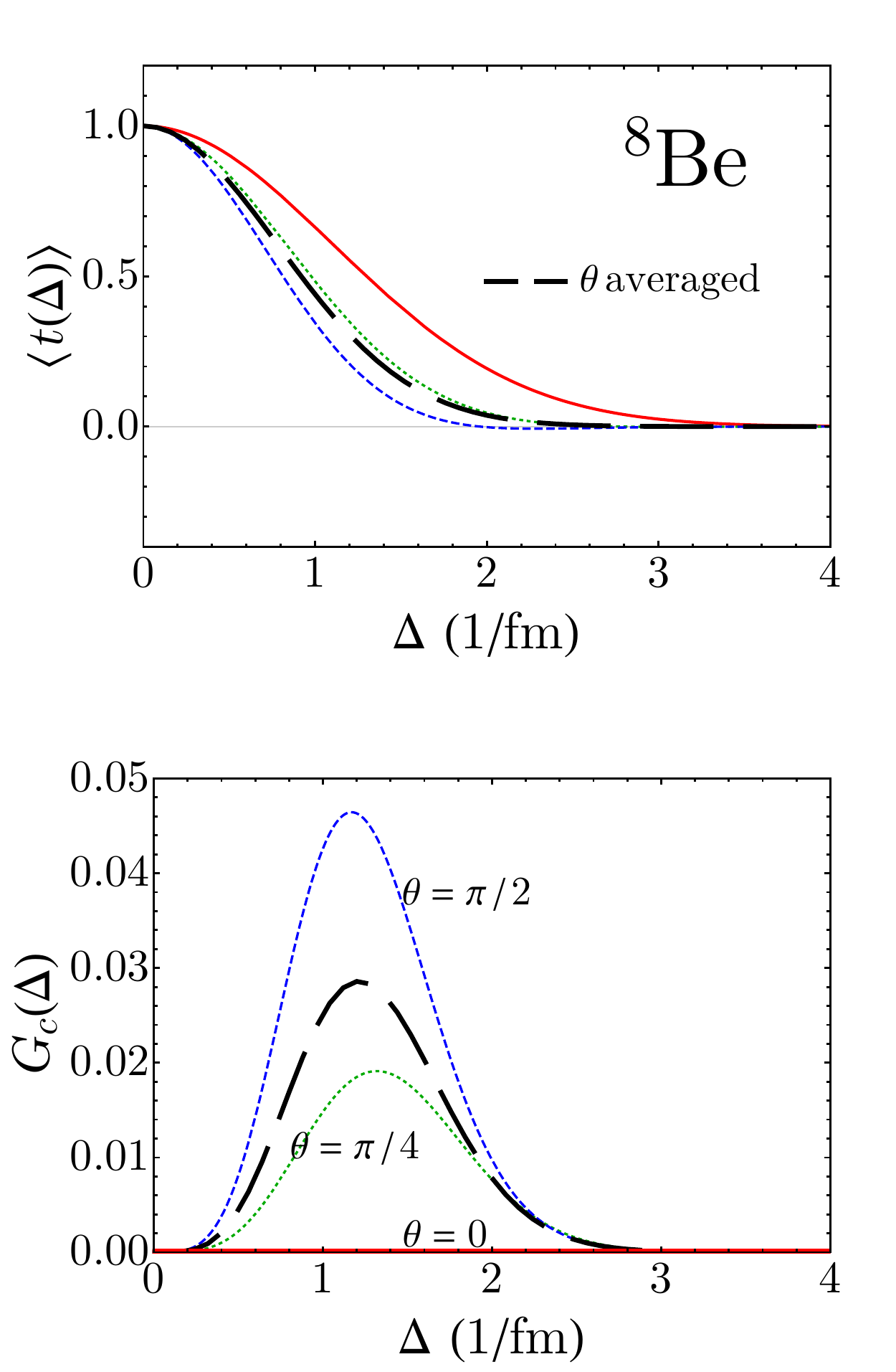}
\caption{Top: Fourier transform, $\langle t(\Delta) \rangle\equiv \langle t_\theta (\Delta) \rangle$ of the thickness function $\langle t_\theta(r) \rangle$ for various values of $\theta$. Bottom: same but for the two-body connected correlation function $G_c(\Delta)$. The black dashed lines represent results averaged over $\theta$. Note that $\langle t_\theta(\Delta=0)\rangle=1$ is expected since $\langle  t_\theta(\b)\rangle$ is normalized to unity.}
\label{fig:FT8Be}
\end{center}
\end{figure}

The Fourier transform of $G(\Delta)$ is given by 
\beq\label{eq:tofDelta}
&&G(\theta,\Delta)=\frac{1}{64} \rme^{ \frac{\Delta^2}{4}\left(\frac{1}{b_1}+\frac{1}{b_2}  \right)}\nn
&&\times \left\{ \! \left( 64 \!-\!16 \frac{\Delta_\theta^2}{\alpha_z^2} \!+\!\frac{\Delta_\theta^4}{\alpha_z^4}\right) I_0(a)\!+\!\left( \!-16 \frac{\Delta_\theta^2}{\alpha_z^2}+2\frac{\Delta_\theta^4}{\alpha_z^4} \right) I_1(a)\right. \nn && \left.+\frac{\Delta_\theta^4}{2 \alpha_z^4}\left[(I_0(a)+I_2(a)\right] \right\},
\eeq
where here $a \equiv \frac{\Delta^2}{4}\left(\frac{1}{b_1}-\frac{1}{b_2}  \right)$. From this equation and that giving $\langle t(\Delta)\rangle$, Eq.~(\ref{eq:DeltathetaBe}), one obtains easily the connected correlation $G_c(\Delta)$. This is plotted in the lower panel of Fig.~\ref{fig:FT8Be}, for various values of the angle $\theta$.  As expected, the correlations are maximal for $\theta=\pi/2$ (nucleus rotating around an axis perpendicular to its symmetry axis) and are absent for $\theta=0$ (in which case there is no collective rotation, the thickness function being then invariant under rotation).

\subsection{The diffractive cross section}

\begin{figure}[t]
    \centering
    \includegraphics[width=.98\linewidth]{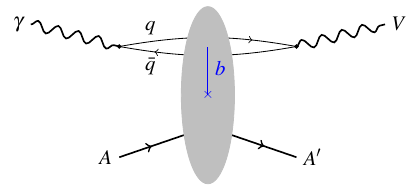}
    \caption{Diffractive production of a vector meson, $V$, in a high-energy $\gamma A$ interaction mediated by a $q \bar q$ dipole.}
    \label{fig:tikz}
\end{figure}

We now turn our attention to a physical process realized in experiments, that of the diffractive electro- or photo-production of vector mesons off a nucleus, illustrated in Fig.~\ref{fig:tikz}. Several steps are involved in the process.  First, the incoming (possibly virtual) photon fluctuates into a pair of nearly on-shell quark and antiquark, forming a small color singlet dipole\footnote{The size of the dipole is determined by the wave function of the vector meson, and in case the vector meson is made of heavy quarks, such as for the J/$\Psi$ meson, is typically much smaller than the size of a nucleon.} that subsequently interacts with one nucleon in the target nucleus (we ignore here multiple scattering, or saturation effects). In this interaction, the transverse size and orientation of the color dipole can be considered as frozen, as are the positions of the nucleons in the nucleus (see e.g.  \cite{Caldwell:2010zza}). Finally, the scattered dipole recombines to form the vector meson.

The amplitude of the process can be shown to be proportional to a matrix element of the thickness function \cite{Caldwell:2010zza}, seen as a one-body operator acting on the transverse coordinates of the nucleon, $\hat t(\x)\!=\!\frac{1}{A}\sum_{i=1}^A\delta(\x\!-\!\hat\b_i)$. Thus, the \textit{elastic} amplitude is proportional to the average of the Fourier transform of this operator in the ground state of the nucleus. Taking into account the form factor of the proton, the elastic amplitude reads 
\beq
\bra{\Psi_0}{\cal A}(\Delta)\ket{\Psi_0}=A\sigma_{\rm dip} \,t_G(\Delta) \,\bra{\Psi_0}\hat t(\Delta)\ket{\Psi_0},
\eeq
where $t_G(\Delta)=\rme^{-\frac{1}{2} B_G \Delta^2}$ is the proton form factor, with typically $B_G=4 \text{GeV}^2$, and where $\sigma_{\rm dip}$ is a (suitably averaged) dipole-nucleon total cross section. In the expression above, $\Delta$, the variable conjugate to the impact parameter, is the momentum transferred to the nucleus in its interaction with the dipole.

At a given momentum transfer $\Delta$, the \textit{coherent} diffractive cross section is proportional to the square of the elastic amplitude, that is
\beq\label{coherentSigTA}
|\bra{\Psi_0}{\cal A}(\Delta)\ket{\Psi_0}|^2= A^2 \sigma_{\rm dip}^2 \,t_G(\Delta)^2 \,|\bra{\Psi_0}\hat t(\Delta)\ket{\Psi_0}|^2
\eeq
The elastic amplitude assumes that only the ground state of the nucleus contributes as an intermediate state. By relaxing this assumption and allowing for all possible intermediate states $\ket{\psi_n}$, one obtains the total diffractive cross section as 
\beq \label{averageSig2}
A^2 \sigma_{\rm dip}^2 t_G(\Delta)^2 \sum_n \left\vert  \bra{\Psi_n}\hat t(\Delta) \ket{\Psi_0}\right\vert^2=\langle {\cal A}(\Delta)^2\rangle.
\eeq 
By combining with Eq.~(\ref{coherentSigTA}), one finds that the \textit{incoherent} cross section (in the considered weak field limit) is proportional to 
\beq
\langle {\cal A}(\Delta)^2\rangle-\langle {\cal A}(\Delta)\rangle^2 =\sigma_{\rm dip}^2\;t_G(\Delta)^2 S(\Delta),
\eeq
where $S(\Delta)$ is given in Eq.~(\ref{eq:SDelta2}). In this equation, 
the angular brackets denote the average over configurations of nucleons in the nucleus, that is, over the ground state of the nucleus. It is convenient to isolate the contribution of the term that is constant in $S(\Delta)$, and set
\beq
\sigma_0= A  \sigma_{\text{dip}}^2 \, \left[ t_G(\Delta)\right]^2,
\eeq
which represents the cross section for independent scattering off $A$ uncorrelated nucleons, and which dominates the cross section at large $\Delta$. 
With this definition, one can express the coherent, total and incoherent diffractive cross sections, respectively, as follows
\beq
\label{eq:dSdeco}
&&\frac{\sigma_{\text{coh}}(\Delta)}{\sigma_0}= A \langle t(\Delta)\rangle^2,\nn 
&&\frac{\sigma_{\text{tot}}}{\sigma_0}= 1+ (A-1) G(\Delta),\nn
&&\frac{\sigma_{\text{inc}}}{\sigma_0}= \frac{\sigma_{\rm tot}-\sigma_{\rm coh}}{\sigma_0} = 1-G(\Delta) +A G_c(\Delta).
\eeq
Note that 
\beq
\frac{\sigma_{\text{tot}} - \sigma_{\text{inc}} }{\sigma_0}= \frac{\sigma_{\text{coh}}(\Delta)}{\sigma_0}.
\eeq

\begin{figure}[t]
\begin{center}
\includegraphics[width=\linewidth]{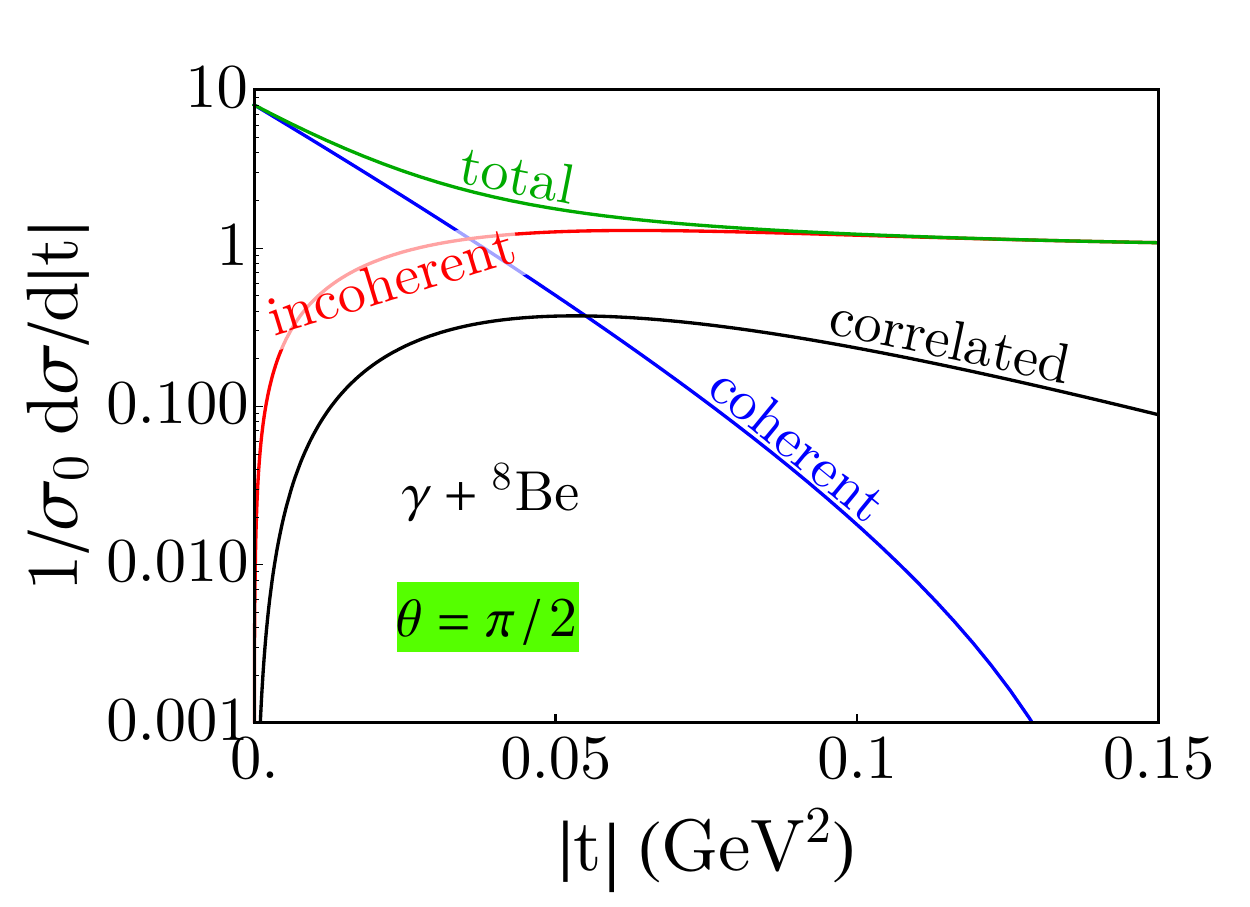}
\vspace{0cm}
\caption{Decomposition of the total diffractive cross section (green line) for the production of a vector meson in $\gamma+^{8}Be$ collisions, as a function of the momentum transfer $|t|$, for $\theta=\pi/2$. The leading correlated part, shown as a black line, contributing to the incoherent cross section (red line) corresponds to $AG_c({\Delta})$ in Eq.~(\ref{eq:dSdeco}).}
\label{fig:8Be_cs}
\end{center}
\end{figure}

The various contributions to the total diffractive cross section are shown in Fig.~\ref{fig:8Be_cs} as a function of $|t|\equiv -\Delta^2$. The plot is made for the case where the beryllium nucleus is rotating in the transverse plane ($\theta=\pi/2$) to render the angular correlation maximal. As anticipated, the total cross section is dominated by coherent diffraction for small $|t|$, and by the incoherent contribution beyond the diffraction minimum. The correlated part $AG_c(|t|)$, the leading contribution of the angular correlations, is most significant in the vicinity of the diffractive minimum whose location is controlled by the size of the nucleus. As illustrated in Fig.~\ref{fig:8BeGG} this corresponds indeed to the scale at which the impact of the nuclear deformation is maximal. We note that this qualitative behavior is consistent with that observed by M\"antysaari \textit{et al.} for $\gamma+^{238}$U collision simulations \cite{Mantysaari:2023qsq}.

\begin{figure}[t]
    \centering
    \includegraphics[width=\linewidth]{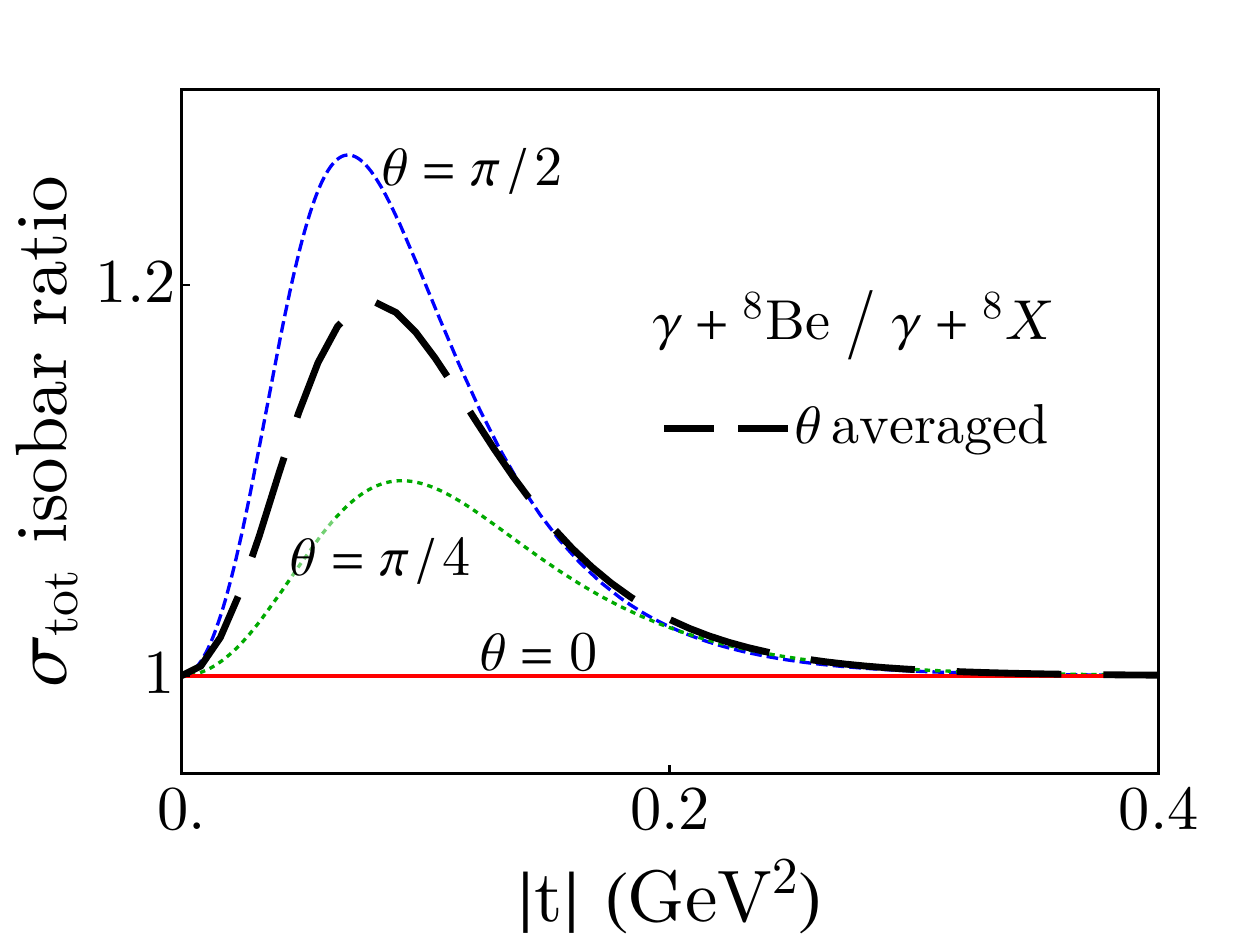}
    \caption{Ratio (\ref{eq:ratioBeX}) of total cross sections, $\sigma_{\rm tot}$, differential in $|t|$, calculated between $\gamma+^{8}$Be and $\gamma+^8$X collisions, where $^8$X is a fictitious isobar of beryllium-8 that is spherical and shares the same thickness function. Different line styles correspond to different choices of the angle $\theta$.}
    \label{fig:isobar}
\end{figure}

\subsection{Isolating nuclear deformation effects with isobars}

To conclude our analysis, we comment on \textit{isobar} collisions as an ideal experimental setup to detect the effects of the nuclear deformation in $\gamma A$ collisions. Let us assume that there exists a fictitious isobar of $^{8}$Be, say $^8$X, whose spherical intrinsic density matches that of $^{8}$Be after averaging over orientations. For such a nucleus, angular correlations are absent, which implies that $G(\Delta)=\langle t(\Delta) \rangle^2$. If vector mesons from diffractive events were collected in these isobaric systems, then the effect of the deformation of $^{8}$Be could be conveniently extracted from the ratio of total cross sections, that is,
\beq\label{eq:ratioBeX}
\frac{\sigma_{\rm tot}\,[\gamma+^{8}{\rm Be}]}{\sigma_{\rm tot}\,[\gamma+^{8}{\rm X}]}=\frac{1+(A-1)G(|t|)}{1+(A-1)\langle  t(|t|) \rangle^2}.
\eeq
Note that this ratio is independent of $\sigma_0$, meaning that several theoretical uncertainties cancel, such as those related to the precise modeling of the diffractive amplitude, the dipole cross section, the vector meson wave function, the form factor of the nucleons at high energy. The ratio (\ref{eq:ratioBeX}) is shown in Fig.~\ref{fig:isobar} for different orientations ($\theta$) of the nucleus. The ratio is unity for small $|t|$ since $\langle \tilde t(|t|=0) \rangle = G(|t|=0)=1$. It also goes to unity when $|t|$ is large. In agreement with our previous analysis, the ratio is significant only in an intermediate range of $|t|$ in the vicinity of the diffraction minimum, where the deformation of $^{8}$Be induces a substantial correction to the total cross section. In our results in Fig.~\ref{fig:isobar}, after averaging over all orientations we observe indeed a maximal 20\% enhancement of the total cross section, localized in the vicinity of the diffractive minimum of the coherent curve ($|t|\approx0.1$ GeV$^2$).

Obviously, for $A=8$ such a pair of isobar nuclei does not exist. However, it is noteworthy that some experimental results for the proposed ratio in isobar collisions, performed with $A=96$ Ru and Zr isotopes, are already available from ultra-peripheral Ru+Ru and Zr+Zr collisions at the Relativistic Heavy Ion Collider \cite{Zhao:2023}. These results are also compared to theoretical calculations in Ref.~\cite{Mantysaari:2023prg}, where a significant enhancement of the total cross section in $\gamma+^{96}$Ru collisions is predicted, compared to the case of $\gamma+^{96}$Zr collisions, as a consequence of the larger quadrupole deformation of $^{96}$Ru. The peak of the enhancement is observed precisely for values of $|t|$ close to the location of the diffraction minimum of the coherent curves. The analysis performed in this paper confirms this prediction, and clarifies its origin. From these considerations, we expect that the final analysis of the STAR results on isobar collisions will give clear evidence of the presence of a peak in the cross section ratio driven by the quadrupole deformation of $^{96}$Ru.

\section{Conclusion and Outlook}
\label{sec:5}

High-energy scattering processes have opened a new avenue for studying the many-body properties of nuclear ground states. Because the time scales involved in these processes are so small as compared to those involved in the intrinsic dynamics of the nuclei, these high-energy processes allow us to probe instantaneous configurations of nucleons thereby revealing their many-body correlations. In particular, the sensitivities of these experiments to the nuclear shape is the underlying motivation of the present work to analyze the characteristic features of angular correlations that are responsible for nuclear deformation.

More specifically, we have used a semi-classical approximation, based on rotor models often used to describe rotational phenomena in nuclear physics, in order to identify characteristic features of angular correlations induced by the average over the orientation of the nucleus. Focusing on axially symmetric systems with quadrupole deformation, we considered various rotor models and showed that the quadrupole deformation leads to a modulation of the azimuthal two-body correlations, generically proportional to $\cos (2\varphi_{12})$ for small deformation, whose magnitude is amplified at the edges of the nuclear system. Although we have considered mainly two-body correlations, angular correlations of a similar nature are expected to affect all $n$-point correlation functions. These could be analyzed following the same strategy as used in this paper. For instance, higher-order deformations, such as the octupole deformation, which has been detected in high-energy collisions \cite{STAR:2021mii,Zhang:2021kxj}, or the triaxiality (also revealed at high energies \cite{Bally:2021qys,ATLAS:2022dov}), would lead to characteristic angular modulation of the nuclear three-body density \cite{Giacalone:2023hwk}.

Although our analysis is based on simplifying assumptions, we expect that the main features of the angular correlations that we have identified would be present in a more refined microscopic treatment. 
 It would indeed be interesting to investigate in more refined theoretical calculations the interplay of angular correlations with other kind of correlations, such as those induced by short-range potentials, or the Pauli exclusion principle, or the center of mass motion. To our knowledge, such a detailed systematic theoretical analysis of the spatial structure of long-range correlations in deformed nuclei across a wide set of isotopes has never been performed.

As a specific example where angular correlations could be identified experimentally, we have considered the diffractive production of vector mesons in $\gamma A$ scattering. We have seen that the cross section of the process gives direct access to the Fourier transform of the density-density correlation function of the nucleus ground state. For a nucleus with a sizable deformation like $^{8}$Be, the effect of the deformation on the total cross section is quite significant, producing a 20\% correction in the proximity of the diffractive minimum. In an upcoming study, we will present results of calculations for a wider set of more accessible nuclear species, including in particular $^{16}$O and $^{20}$Ne, which are of direct relevance for upcoming experimental campaigns \cite{Mariani:2024jzp,AlemanyFernandez:2025ixd}. Although these nuclei are not isobars, the same technique based on a ratio of total cross sections, such as that in Eq.~(\ref{eq:ratioBeX}), could be used in the analysis of ultra-peripheral collisions (see e.g. \cite{Mantysaari:2023qsq}) to reveal, in particular, the elongated geometry of $^{20}$Ne. The detailed impact of nuclear clustering on the cross sections deserves as well a thorough investigation \cite{Magdy:2024thf}.

As an intriguing outlook, we note that the formula for the diffractive cross section bears a resemblance with the so-called nuclear matrix element (NME) entering the conjectured neutrinoless double beta decay ($0\nu\beta\beta$) transition \cite{Agostini:2022zub}, whose knowledge is crucial in the design of future experimental searches. In this process, two neutrons are converted into two protons. For the spatial part, the NME involves the Fourier transform of the expectation value of a two-body operator \cite{Yao:2021wst}, in a strong analogy with our Eq.~(\ref{eq:SDelta}). In fact, the NME is highly sensitive to nuclear quadrupole deformation whenever the parent nucleus is well deformed \cite{Zhang:2024tzr,Li:2025vdp}. As $0\nu\beta\beta$ decay occurs between isobaric species, we expect that an analysis of the \textit{isobar ratio} of cross sections discussed in Fig.~\ref{fig:isobar} for these isotopes should yield detailed information about the NME.

All in all, a program focused on understanding the interplay of many-body nuclear phenomena that are central in nuclear structure research, such as nuclear deformations or the NME of $0\nu\beta\beta$ decay, and the observables accessible in high-energy scattering experiments, is only just beginning \cite{Duguet:2025hwi}. We anticipate significant developments along these lines in the near future.

\begin{acknowledgements}
We thank Gordon Baym, Jiangming Yao, and Wenbin Zhao for useful discussions. We acknowledge the hospitality of the China Center of Advanced Science And Technology (CCAST, Beijing) in the context of the program+workshop \textit{``Exploring nuclear physics across energy scales 2024: intersection between nuclear structure and high-energy nuclear collisions''}, where this work was initiated.
\end{acknowledgements}

%\bibliographystyle{unsrt}
%\bibliography{biblio}

\end{document}